\pdfoutput=1
\documentclass[cernpreprint=true,11pt,a4paper,UKenglish,texlive=2013,texmf]{article}
\usepackage{jheppub}
\usepackage[biblatex=false,default=false,minimal=true]{atlaspackage}
\usepackage{xspace}
\usepackage{atlasphysics}
\newcommand{\Npart}{\mbox{$N_{\mathrm{part}}$}\xspace}

\newcommand{\avgNpart}{\mbox{$\langle \Npart \rangle$}\xspace}
\newcommand{\Ncoll}{\mbox{$N_{\mathrm{coll}}$}\xspace}

\newcommand{\Taa}{\mbox{$T_{\mathrm{AA}}$}\xspace}
\newcommand{\TaaPerif}{\mbox{$T_{\mathrm{AA}}^{60-80\%}$}\xspace}

\newcommand{\avgTaa}{\mbox{$\langle \Taa \rangle$}\xspace}
\newcommand{\avgTaaPerif}{\mbox{$\langle \TaaPerif \rangle$}\xspace}

\newcommand{\dsigmadeta}{\mbox{$\mathrm{d}\sigma/\mathrm{d}\eta$}\xspace}

\newcommand{\Nevt}{\mbox{$N_{\mathrm{evt}}$}\xspace}

\newcommand{\energy}{\mbox{$\sqrt{s_{\mathrm{NN}}}=2.76$\,\TeV}\xspace}
\newcommand{\sqn}{\mbox{$\sqrt{s_{\mathrm{NN}}}$}\xspace}
\newcommand{\sqs}{\mbox{$\sqrt{s}$}\xspace}

\newcommand{\PbPb}{\mbox{Pb+Pb}\xspace}

\newcommand{\pA}{\mbox{$p$+A}\xspace}
\newcommand{\aA}{\mbox{AA}\xspace}
\newcommand{\pp}{\mbox{$pp$}\xspace}

\newcommand{\rcp}{\mbox{$R_{\mathrm{CP}}$}\xspace}

\newcommand{\raa}{\mbox{$R_{\mathrm{AA}}$}\xspace}
\newcommand{\rpbpb}{\mbox{$R_{\mathrm{AA}}$}\xspace}

\newcommand{\zz}{\mbox{$z_0\sin\theta$}}
\newcommand{\dz}{\mbox{$d_0$}}
\newcommand{\edz}{\mbox{$\sigma_{\dz}$}}
\newcommand{\ezz}{\mbox{$\sigma_{\zz}$}}

\newcommand{\akt}{\mbox{anti-$k_{t}$}\xspace}

\hypersetup{pdftitle={ATLAS draft},pdfauthor={The ATLAS Collaboration}}
\usepackage[coverpage=false]{atlascover}
\newcommand{\figdir}{.}
\newcommand{\papertype}{paper\xspace}

\AtlasTitle{Measurement of charged-particle spectra in Pb+Pb collisions at \mbox{$\sqrt{{s}_\mathsf{NN}} = {}$2.76\,Te\kern -0.05em V} with the ATLAS detector at the LHC} 

\AtlasJournal{JHEP}

\AtlasAbstract{

Charged-particle spectra obtained in \PbPb interactions at $\sqn=2.76$\,\TeV\ and \pp interactions at $\sqs=2.76$\,\TeV\ with the ATLAS detector at the LHC are presented, using data with integrated luminosities of 0.15\,nb${}^{-1}$ and 4.2\,pb${}^{-1}$, respectively, in a wide transverse momentum ($0.5 < \pT < 150$\,\GeV) and pseudorapidity ($|\eta|<2$) range. For \PbPb collisions, the spectra are presented as a function of collision centrality, which is determined by the response of the forward calorimeters located on both sides of the interaction point. The nuclear modification factors $\raa$ and $\rcp$ are presented in detail as a function of centrality, \pT\ and $\eta$. They show a distinct \pT-dependence with a pronounced minimum at about 7\,\GeV. Above 60\,\GeV, $\raa$ is consistent with a plateau at a centrality-dependent value, within the uncertainties. The value is \mbox{$0.55\pm0.01(stat.)\pm0.04(syst.)$} in the most central collisions. The $\raa$ distribution is consistent with flat $|\eta|$ dependence over the whole transverse momentum range in all centrality classes.
}

\PreprintIdNumber{CERN-PH-EP-2015-032}

\begin{document}

\title{\AtlasTitleText}

\author{The ATLAS Collaboration}

\begin{abstract}
	\AtlasAbstractText
\end{abstract}

\maketitle

\section{Introduction} 
\label{sec:intro}

High-energy heavy-ion (HI) collisions produce a hot and dense matter, the quark-gluon plasma, in which quarks and gluons become deconfined~\cite{brahms,phobos,star,phenix}. Charged hadrons of high transverse momentum (\pT) are a tool that can be used to study this strongly interacting matter~\cite{denterria,solana}. Since the start of the LHC operation with heavy ions in 2010, it has been possible to study charged-particle production in HI collisions in a new energy regime.

The first results from the LHC experiments showed that jets emerging from the medium created in such collisions have lower energy than expected in the absence of medium effects~\cite{Aj_atlas,Aj_cms}. The measurement of fully reconstructed jets by the ATLAS experiment revealed that the energy loss of the high-energy partons in the medium, commonly referred to as jet quenching, results in a lower yield of jets at fixed \pT. The yield is suppressed in central collisions with respect to the yield in \pp collisions by factor of two to four~\cite{atlas_jet_raa,alice_jet_raa}. Additional important information about the energy-loss mechanism is provided by the study of jet fragmentation functions~\cite{frag_atlas,frag_cms}. These show an enhancement in the charged-particle fragment yield in central collisions with respect to those measured in the peripheral ones for transverse momentum $\pT \lesssim 4$\,\GeV, a reduction for 4\,\GeV${} \lesssim \pT \lesssim 30$\,\GeV\ and a small enhancement for $\pT \gtrsim 30$\,\GeV.

As jets lose energy and the pattern of jet fragmentation is modified in HI collisions, it is expected that the resulting spectrum of hadrons, which originate from parton fragmentation, is also modified. The results from the LHC experiments~\cite{cms_raa, alice_raa} established that hadron suppression at the LHC is larger than that measured at RHIC~\mbox{\cite{brahms,phobos,star,phenix}}. At \pT\ values of 6--7\,\GeV\ the suppression reaches a factor of five for nucleus--nucleus collisions at a centre-of-mass energy per nucleon--nucleon pair $\sqn=200$\,\GeV\ and a factor of seven at $\sqn=2.76$\,\TeV. Results from the LHC experiments show that the suppression diminishes at higher \pT.

Several theoretical models attempt to describe the behaviour of the data at high $\pT$~\mbox{\cite{theo_new_1,theo_new_2}}, using previous measurements by ALICE and CMS; however the behaviour above \mbox{$\pT \approx 50$\,\GeV} is not strongly constrained by the data. An attempt was made to predict the behaviour of the suppression at a fixed \pT\ as a function of rapidity~\cite{theo_Renk}, but such dependence has not yet been measured. The ATLAS tracking system has a wide rapidity coverage and thus allows to measure this pseudorapidity dependence.

The suppression of hadron production in HI collisions can be quantified using the nuclear modification factor (\raa), a ratio of the measured charged-particle production yield in HI collisions to the expected rate based on the charged-particle production in nucleon--nucleon collisions. Collisions with complete nuclear overlap are called central collisions, and as the distance between nuclear centres increases, the collisions become more and more peripheral. The hard scattering rate is higher in more central collisions because of greater nuclear overlap where more nucleons in one nucleus can hard-scatter with nucleons in the other. Such an enhancement is expected by the Glauber model to be proportional to the product of nucleon densities in the transverse plane; the ratio is corrected for this by $\avgTaa$, which is estimated as the number of nucleon--nucleon collisions over their cross section~\cite{Glauber}. Neglecting isospin effects, the nucleon--nucleon hadron-production cross section can be approximated by the corresponding \pp cross section, $\sigma_{\mathrm{NN}}\approx\sigma_{\pp}$, and thus the nuclear modification factor can be expressed as 

\begin{equation}
\rpbpb = \frac{1}{\avgTaa} \frac{1/\Nevt~\mathrm{d}^2N_{\footnotesize\PbPb} / \mathrm{d}\eta \mathrm{d}\pT}{\mathrm{d}^2\mathrm{\sigma}_{\footnotesize\pp} / \mathrm{d}\eta \mathrm{d}\pT},
  \label{eq:raa}
\end{equation}
where $\Nevt$ is the number of \PbPb events, $\mathrm{d}^2N_{\footnotesize\PbPb} / \mathrm{d}\eta \mathrm{d}\pT$ is the differential yield of charged particles in \PbPb collisions and $\mathrm{d}^2\mathrm{\sigma}_{\footnotesize\pp} / \mathrm{d}\eta \mathrm{d}\pT$ is the differential charged-particle production cross section measured in \pp collisions.  It is expected that in the absence of nuclear effects, the ratio \rpbpb will be unity in the region of \pT\ where hadron production is dominated by hard scattering processes.

The modifications of the particle spectra can also be studied using the nuclear modification factor $\rcp$, defined as the ratio of the charged-particle yield in each centrality class to the yield in a chosen peripheral class, both scaled by their corresponding \avgTaa. The nuclear modification factor $\rcp$ is defined as 

\begin{equation}
\rcp(\pT,\eta) =
\frac{\langle T_{\footnotesize{\mathrm{\aA,P}}}\rangle}
       {\langle T_{\footnotesize{\mathrm{\aA,C}}}\rangle} 
\frac{1/N_{\mathrm{evt,C}}~\mathrm{d}^2 N_{\footnotesize{\PbPb,\mathrm{C}}} / \mathrm{d}\eta \mathrm{d}\pT}
       {1/N_{\mathrm{evt,P}}~\mathrm{d}^2 N_{\footnotesize{\PbPb,\mathrm{P}}} / \mathrm{d}\eta \mathrm{d}\pT},
\label{eq:rcp}
\end{equation}

\noindent where indices P and C denote quantities obtained in peripheral collisions and for a given centrality, respectively. Measuring \rcp\ is another way to examine the suppression of hadron production in HI collisions.

This \papertype presents a high-statistics measurement of charged-hadron spectra, \rpbpb and \rcp\ in the transverse momentum range $0.5<\pT<150$\,\GeV\ in \PbPb\ collisions using data recorded by the ATLAS experiment in the 2010 and 2011 HI physics runs of the LHC at $\sqn=2.76$\,\TeV, and the \pp data at the same value of \sqs recorded in the 2011 and 2013 physics runs. The total integrated luminosity of the combined \PbPb sample is 0.15\,nb${}^{-1}$, and that of the combined \pp sample is 4.2\,pb${}^{-1}$.

\section{The ATLAS detector} 
\label{sec:detector}

The measurements presented in this \papertype were performed using the ATLAS inner detector~(ID), calorimeter, muon spectrometer, and the high-level trigger and data acquisition systems \cite{Aad:2008zzm}. The ID measures charged-particle tracks within the pseudorapidity\footnote{ATLAS uses a right-handed coordinate system with its origin at the nominal interaction point (IP) in the centre of the detector and the $z$-axis along the beam pipe. The $x$-axis points from the IP to the centre of the LHC ring, and the $y$-axis points upward. Cylindrical coordinates $(r,\phi)$ are used in the transverse plane, $\phi$ being the azimuthal angle around the beam pipe. The pseudorapidity is defined in terms of the polar angle $\theta$ as $\eta=-\ln\tan(\theta/2)$.} interval $|\eta|<2.5$ using a combination of silicon pixel detectors (Pixel), silicon microstrip detectors (SCT), and a straw tube transition radiation tracker (TRT), all immersed inside a 2\,T axial magnetic field. All three tracker systems are composed of a barrel and two symmetrically placed end-cap sections. The Pixel is composed of three layers of sensors with nominal pixel size $50\,{\rm \mu m} \times 400\,{\rm \mu m}$. The SCT barrel section contains four layers of double-sided modules with 80\,$\mu {\rm m}$ pitch sensors, while the end-caps consist of nine layers of double-sided modules with radial strips having a mean pitch of $80~\mathrm{\mu m}$. The two sides of each layer in both the barrel and the end-caps have a relative stereo angle of 40\,mrad. The TRT contains up to 73 (160) layers of staggered straws interleaved with fibers (foils) in the barrel (end-cap). Charged particles in the barrel region with $\pT \gtrsim 0.5$~\GeV\ and $|\eta| < 1.0$ typically traverse 3 layers of silicon pixel detectors, 4 layers of double-sided microstrip modules, and 36 straws.

The calorimeter system consists of a liquid argon (LAr) electromagnetic (EM) calorimeter covering $|\eta|<3.2$, a steel--scintillator sampling hadronic calorimeter covering $|\eta| <1.7$, two LAr hadronic calorimeters covering $1.5 < |\eta| < 3.2$, and two LAr forward calorimeters (FCal) covering $3.1 < |\eta| < 4.9$. The two FCal modules are composed of tungsten and copper absorbers with LAr as the active medium, which together provide ten interaction lengths of material. The hadronic calorimeter has three sampling layers longitudinal in shower depth in $|\eta|<1.7$ and four sampling layers in $1.5<|\eta|<3.2$, with a slight overlap of the two regions. It has a $\Delta \eta \times \Delta \phi$ granularity of $0.1 \times 0.1$ for $|\eta| < 2.5$ and  $0.2 \times 0.2$ for $2.5 < |\eta| < 4.9$.\footnote{An exception is the third sampling layer which has a segmentation of $0.2 \times 0.1$ up to $|\eta| = 1.7$.} The EM calorimeter is segmented longitudinally in shower depth into three compartments with an additional pre-sampler layer. It has a granularity that varies with layer and pseudorapidity, but which is generally much finer than that of the hadronic calorimeter. The middle sampling layer, in which EM showers typically deposit the largest fraction of their energy, has a granularity of $0.025 \times 0.025$ over $|\eta| < 2.5$. 

The minimum-bias trigger scintillator (MBTS) counters are located at a distance of 3.56\,m along the beamline from the centre of the ATLAS detector and cover $2.1<|\eta|<3.9$ on each side. Each MBTS counter consists of 16 scintillator pads. Signals from the pads were used as an input to the trigger system in the 2010 run and the time difference between hits in the two sides of the MBTS was used to select good events in both the 2010 and 2011 runs.

The zero degree calorimeter (ZDC) consists of two arms, positioned at $z = \pm140$\,m from the centre of the ATLAS detector, and detects neutrons and photons with $|\eta|>8.3$. Signals from the ZDC are used by the trigger systems. The ZDC trigger thresholds were set just below the single-neutron peak on each side.

The selection of events was done in several steps. First, events were required to satisfy a hardware-based level-1 (L1) trigger. The L1 trigger selects events with energy deposition in the calorimeters above a preset level, or events with signals from the ZDC. Such events are further processed by software-based high-level triggers (HLT).

\section{Datasets}
\label{sec:datasets}

This analysis uses data from the 2010 and 2011 \PbPb data-taking at $\sqn=2.76$\,\TeV\ and data from the 2011 and 2013 \pp data-taking at the same centre-of-mass energy, \mbox{$\sqs=2.76$\,\TeV}. A summary of all data samples used in this analysis is given in table~\ref{table:samples}. 

\begin{table*}[htb]
\begin{center}
\begin{tabular}{|c|ccccc|} \hline
 \multicolumn{6}{|c|}{\PbPb data} \\ \hline
 Year  & Trigger & Centrality & Recorded & Sampled  & NN collisions\\ \hline
 2010   & MB     &0--80\%   & $4.2\times10^{7}$   & $4.2\times10^{7}$ & $1.8\times10^{10}$\\
 2011   & MB     &0--80\%   & $4.2\times10^{7}$   & $8.0\times10^{8}$ & $3.4\times10^{11}$\\
 2011   & Jets   &0--80\%   & $1.3\times10^{7}$  & $8.0\times10^{8}$ & $3.4\times10^{11}$\\ 
 2011   & Jets   &0--5\%    & $2.5\times10^{6}$  & $5.0\times10^{7}$ & $8.4\times10^{10}$\\ 
 2011   & Jets   &60--80\%  & $3.3\times10^{6}$  & $2.0\times10^{8}$ & $5.3\times10^{9}$\\ \hline
 \end{tabular} 
 \vspace*{0.3cm}\\
 \begin{tabular}{|c|cccc|}  \hline
 \multicolumn{5}{|c|}{\pp data} \\ \hline
  Year  & Trigger &  Recorded & Luminosity  & Collisions\\ \hline
 2011   & MB    & $1.6\times10^{7}$ & 0.2\,pb${}^{-1}$ & $1.3\times10^{10}$\\
 2011   & Jets  & $6.5\times10^{7}$ & 0.2\,pb${}^{-1}$ & $1.3\times10^{10}$\\
 2013   & MB    & $5.5\times10^{6}$ & 4.0\,pb${}^{-1}$ & $2.6\times10^{11}$\\ 
 2013   & Jets  & $6.8\times10^{5}$ & 4.0\,pb${}^{-1}$ & $2.6\times10^{11}$\\  \hline
 \end{tabular}
 \vspace*{0.3cm}\\
 \begin{tabular}{|c|ccc|} \hline
 \multicolumn{4}{|c|}{\PbPb simulation samples} \\  \hline
Year  & Underlying event &  Hard scattering & Total events\\ \hline
 2010 & HIJING         &    PYTHIA &  $5\times10^{6}$\\
 2011 & Data   &  PYTHIA &  $2.2\times10^{7}$\\ \hline
 \end{tabular}
 \vspace*{0.3cm}\\
 \begin{tabular}{|c|cc|} \hline
 \multicolumn{3}{|c|}{\pp simulation samples} \\  \hline
Year  &   Hard scattering & Total events\\ \hline 
 2011  &  PYTHIA &  $3.5\times10^{7}$\\
 2013  &  PYTHIA &  $1.9\times10^{7}$\\ \hline
\end{tabular}
\caption{Summary of the data and simulation samples used in the analysis. ``MB" stands for minimum bias. For \PbPb data, the column ``Recorded" lists how many events for a given data-taking period, trigger and centrality interval were written to disk. ``Sampled" is the total number of events, of which only a fraction were recorded. The ``NN collisions" column in \PbPb lists the number of binary nucleon--nucleon collisions equivalent to the ``Sampled" events based on the Glauber model~\cite{Glauber}. For \pp data, the meaning of the column ``Recorded" is analogous. ``Luminosity" is the total luminosity for a given run, while ``Collisions" gives the equivalent number of proton--proton collisions for that luminosity. For the simulation samples, the tables specify the origin of the underlying event (for the \PbPb samples) and the hard scattering event. The last columns sum up the total number of generated events. These events may be further broken down, such as by kinematic selection on the underlying hard scattering (see text).}
\label{table:samples}
\end{center}
\end{table*}

All 2010 \PbPb data were collected with a minimum-bias (MB) trigger. It required a presence of a neutron or photon on either side of the ZDC or presence of a hit in the MBTS on each side of the ATLAS detector.

In the  2011 data-taking, events were recorded with a MB trigger and a jet trigger. The transverse energy measurements used by the trigger system is evaluated at the electromagnetic scale. The MB events were required either to have the transverse energy, \et, in the whole calorimeter exceeding 50\,\GeV\ at the L1 trigger or to have an ID track reconstructed in the event in coincidence with the ZDC signals on both sides. For a jet trigger, events were preselected for reconstruction by the HLT if they deposited a transverse energy of $\et>10$\,\GeV\ in the calorimeters at L1 or if the ZDC produced signals in coincidence on both sides of the detector. Such events were analysed by the HLT using a jet-finding algorithm on calorimeter towers after underlying-event subtraction \cite{jet_paper_new}. All events containing a jet found with the \akt algorithm \cite{anti-kt} with radius parameter $R=0.2$ and with \et, estimated in the trigger system, of at least 20\,\GeV\ were accepted. The HLT uses algorithms very similar to those used in the offline jet reconstruction.

The \pp data with $\sqs=2.76$\,\TeV\ were obtained in two designated runs in 2011 and 2013. In 2011, the MB trigger required a coincidence of MBTS signals. Events selected with the jet trigger in 2011 were required either to satisfy the requirement of MBTS signals in coincidence and a jet with $\et>10$\,\GeV\ or to have an area of $\Delta\eta\times\Delta\phi=0.8\times0.8$ in the calorimeters with deposited transverse energy more than 10\,\GeV\ at L1. 

In 2013, the MB trigger required an event to be randomly selected at L1 in filled bunch crossings and a track to be reconstructed in the ID system. Due to a different trigger menu designed to take advantage of the higher instantaneous luminosity, only a small fraction of MB events was recorded in 2013 compared to 2011. The jet triggers in the 2013 data-taking used different jet \ET\ thresholds. For the jet-triggered samples with the lowest thresholds, jet \mbox{$\et=10$\,\GeV}~and~\mbox{$20$\,\GeV}, events were randomly selected by the L1 trigger and then by the HLT using \akt jet algorithm with a radius parameter of $R=0.4$ and requiring an energy estimate above the corresponding $\ET$ threshold. For the jet-triggered samples with thresholds of $\et=40$\,\GeV~and~$50$\,\GeV, events at L1 were required to deposit transverse energy above 5\,\GeV~and~10\,\GeV, respectively, in an area of $\Delta\eta\times\Delta\phi=0.8\times0.8$ in the ATLAS calorimeters. For the jet-triggered samples with thresholds $\et=60$\,\GeV~and~$75$\,\GeV, events were selected by requiring a transverse energy deposition above 15\,\GeV\ at L1. Due to the limited DAQ bandwidth, the jet-triggered samples were recorded with different prescales depending on the $\ET$ threshold, and these prescales were chosen to evolve with the instantaneous luminosity in the LHC fill. The prescale indicates which fraction of events that passed the triger selection were selected for recording by the data acquisition system.

The integrated luminosities of the \pp\ samples, after application of the event selection criteria, are given in table~\ref{table:samples}. After accounting for the nuclear thickness, the luminosities in \PbPb and \pp samples are comparable, as can be seen in the ``NN collisions" column.

A Monte Carlo (MC) simulation was used to correct the detector-level charged-particle spectrum for track reconstruction efficiency loss, detector resolution and the contribution of tracks not produced by primary particles. The track reconstruction performance was simulated in the HI environment. The particles were generated by the {\sc PYTHIA} event generator~\cite{Pythia} version 6.423 with parameters chosen according to the so-called AUET2B parameter set (``tune")~\cite{pythia_tune}. Five samples of $\sqs = 2.76$~\TeV\ \pp\ hard scattering events were produced in exclusive intervals of transverse momentum of outgoing partons in the $2\rightarrow 2$ hard scattering process, with boundaries at 17, 35, 70, 140, 280 and 560\,\GeV. Each PYTHIA event was embedded in the underlying event of a HI collision, which was obtained using different techniques for the 2010 and 2011 data analyses.

In the 2010 data analysis the environment of \PbPb\ collisions was simulated using $10^{6}$ events, the same for all five samples, with $\sqn = 2.76$~\TeV. They were produced by the HIJING event generator \cite{Wang:1991hta} with default parameters, except that jet quenching was disabled. To simulate the effects of elliptic flow in \PbPb\ collisions, a parameterized centrality-, $\eta$- and \pT-dependent $\cos{2\phi}$ modulation with randomly selected event plane based on previous ATLAS measurements \cite{ell_flow} was imposed on the particles after generation. The detector response in the resulting MC events was evaluated using GEANT4 \cite{Agostinelli:2002hh} configured with geometry and digitization parameters matching those of the 2010 \PbPb\ run \cite{atlas_simul}. 

In the 2011 data-taking a separate \PbPb event sample was recorded which contained approximately $3\times10^{6}$ MB events to be used as the underlying events in the MC samples. Approximately $4.3 \times 10^{6}$ hard scattering PYTHIA events (for each of the five different kinematic selections) were overlaid on these data. Thus, the simulations for the 2011 analysis contains an underlying event contribution identical to the data.

The \pp\ simulated events were generated by PYTHIA, using the same tune as for \PbPb, and reconstructed with the detector conditions matching those from the data-taking period. Minimum-bias and hard scattering samples were simulated for both runs. The MB simulation contained $30\times10^{6}$ and $2\times10^{6}$ events corresponding to 2011 and 2013 runs respectively. The hard scattering samples were produced in several kinematic regions similarly to the HI samples. For the 2013 samples, the kinematic regions were defined using generated leading jet \pT. The total number of events in the hard scattering samples was $5\times10^{6}$ and $17\times10^{6}$ for 2011 and 2013 conditions, respectively.

\section{The centrality of \PbPb collisions}
\label{sec:centr}

\begin{table}[b!]
\begin{center}
\begin{tabular}{|r|c|c|c|} \hline
Centrality& $\langle \Npart \rangle$ & $\langle \Taa \rangle$ [mb$^{-1}$] &{ $\frac{\avgTaa}{\avgTaaPerif}$ }\\ \hline

0--1\%   & 401 $\pm$ 1    & 29.0 $\pm$ 0.5 & 70.2 $\pm$ 8.2 \\
0--5\%   & 382 $\pm$ 2    & 26.3 $\pm$ 0.4 & 63.6 $\pm$ 7.4 \\
5--10\%  & 330 $\pm$ 3    & 20.6 $\pm$ 0.3 & 49.8 $\pm$ 5.6 \\
10--20\% & 261 $\pm$ 4    & 14.4 $\pm$ 0.3 & 34.9 $\pm$ 3.7 \\
20--30\% & 186 $\pm$ 4    & 8.73 $\pm$ 0.26 & 21.1 $\pm$ 2.4 \\
30--40\% & 129 $\pm$ 4    & 5.05 $\pm$ 0.22 & 12.2 $\pm$ 1.0 \\
40--50\% & 85.6 $\pm$ 3.5 & 2.70 $\pm$ 0.17 & 6.5 $\pm$ 0.4 \\
50--60\% & 53.0 $\pm$ 3.5 & 1.34 $\pm$ 0.12 & 3.2 $\pm$ 0.1 \\
60--80\% & 22.6 $\pm$ 2.3 & 0.41 $\pm$ 0.05&  \\ \hline
\end{tabular}
\caption{Centrality classes used in this analysis. The mean number of participants, \avgNpart, the mean value of nuclear overlap function, \avgTaa, and its ratio to the most peripheral \mbox{60--80\%} centrality class as estimated using a Glauber Monte Carlo model are also presented. }
\label{table:centrality}
\end{center}
\end{table}

The centrality of \PbPb collisions is characterized using the total transverse energy ($\sum E_\mathrm{T}$) measured at the electromagnetic scale by the FCal~\cite{ell_flow}. A detailed study based on a Glauber Monte Carlo model \cite{Glauber} using \pp\ data at $\sqs=2.76$\,\TeV\ estimated the fraction of the total Glauber cross section passing the trigger and the event selection cuts used in this analysis as $f=(98\pm2)\%$. Taking into account the visible fraction of the total cross section and using the $\sum E_\mathrm{T}$ distribution, the \PbPb event sample is divided into different centrality classes, defined by the percentiles shown in table~\ref{table:centrality}. For the measurement of $\rcp$ defined in eq.~(\ref{eq:rcp}), the ``peripheral" reference class is chosen to be 60--80\% while the ``central" class refers to the remaining classes of more central collisions.

The mean number of participating nucleons, \avgNpart, as well as \Taa are estimated for each centrality class with the same Glauber calculation. The values are shown in table~\ref{table:centrality}, with uncertainties determined by varying the geometric description, as well as the value of~$f$, used in the Glauber calculation. The last column of table~\ref{table:centrality} shows the ratio of \avgTaa\ in each centrality class to that in the most peripheral (60--80\%) centrality class.

\section{Charged-particle track selection}
\label{sec:track_cuts}
Charged-particle tracks are reconstructed using the ID system in the pseudorapidity region \mbox{$|\eta|<2.5$} and over the full azimuth. The minimum \pT\ of reconstructed tracks in the \PbPb samples is 0.5\,\GeV. The abundance of particles produced in HI collisions results in increased occupancies in the ID subsystems causing a deterioration of the ID tracking performance in the most central HI collisions. The degradation of the tracking performance at high occupancy manifests itself in three different ways: a decrease of the tracking efficiency, a worsening in the track momentum resolution and the reconstruction of fake tracks. Fake tracks are composed of randomly associated hits in the detector layers. These phenomena are discussed in this section.

The analysis of the charged-particle spectra presented in this \papertype refers to primary charged particles directly produced in the nucleus--nucleus interactions with a mean lifetime greater than $0.3\times 10^{-10}$\,s, or long-lived charged particles created by subsequent decays of particles with a shorter lifetime~\cite{ATLASminb}. All other particles are considered secondary. Tracks produced by primary and secondary particles are referred to in what follows as primary and secondary tracks, respectively. An exception is made for muons and electrons coming from the decays of \textit{W} and \textit{Z} bosons, whose contributions are excluded from the results, as these leptons follow binary scaling~\cite{atlas_lept,atlas_lept2,cms_lept,cms_lept2} and they would affect results with different behaviour. Reconstructed tracks arising from a spurious association of detector-layer hits which originate from different particles and result in a reconstructed track are considered to be fake tracks.

In the MC simulation the categorisation of a particle relies on the matching of a reconstructed track to a generated particle. The matching is done based on contributions of generated particles to the hits in the detector layers. A reconstructed track is matched to a generated particle if it contains hits produced primarily by this particle. The matching procedure is explained in ref.~\cite{atlas_simul}. In addition, the reconstructed track \pT\ is required to be within the range from $0.5\pT^{\rm gen}$ to $1.5\pT^{\rm gen}$ to be considered as successfully reconstructed, where $\pT^{\rm gen}$ is the generated transverse momentum of the particle.

\subsection{Modelling of track reconstruction parameters}
\label{sec:track_params}
In HI collisions, the occupancies of the three tracking subsystems increase to different degrees. The Pixel detector occupancy is below 1\% even in the most central collisions. The corresponding number for the SCT detector is below 10\%, while the occupancy in the TRT reaches 90\%. To account for the high occupancy in \PbPb events, the track reconstruction is configured differently from that in \pp collisions~\mbox{\cite{Aad:2010rd,Aad:2010fh}.} In \PbPb reconstruction, a larger weight is assigned to hits in the Pixel and SCT detectors, thus resulting in a procedure more robust against fake track production. All tracks reconstructed in HI collisions are extensions of the track candidates from these two subsystems; TRT-based tracks are not used. 

\begin{figure*}[!tb]
	\centering
	\includegraphics[width=0.95\textwidth]{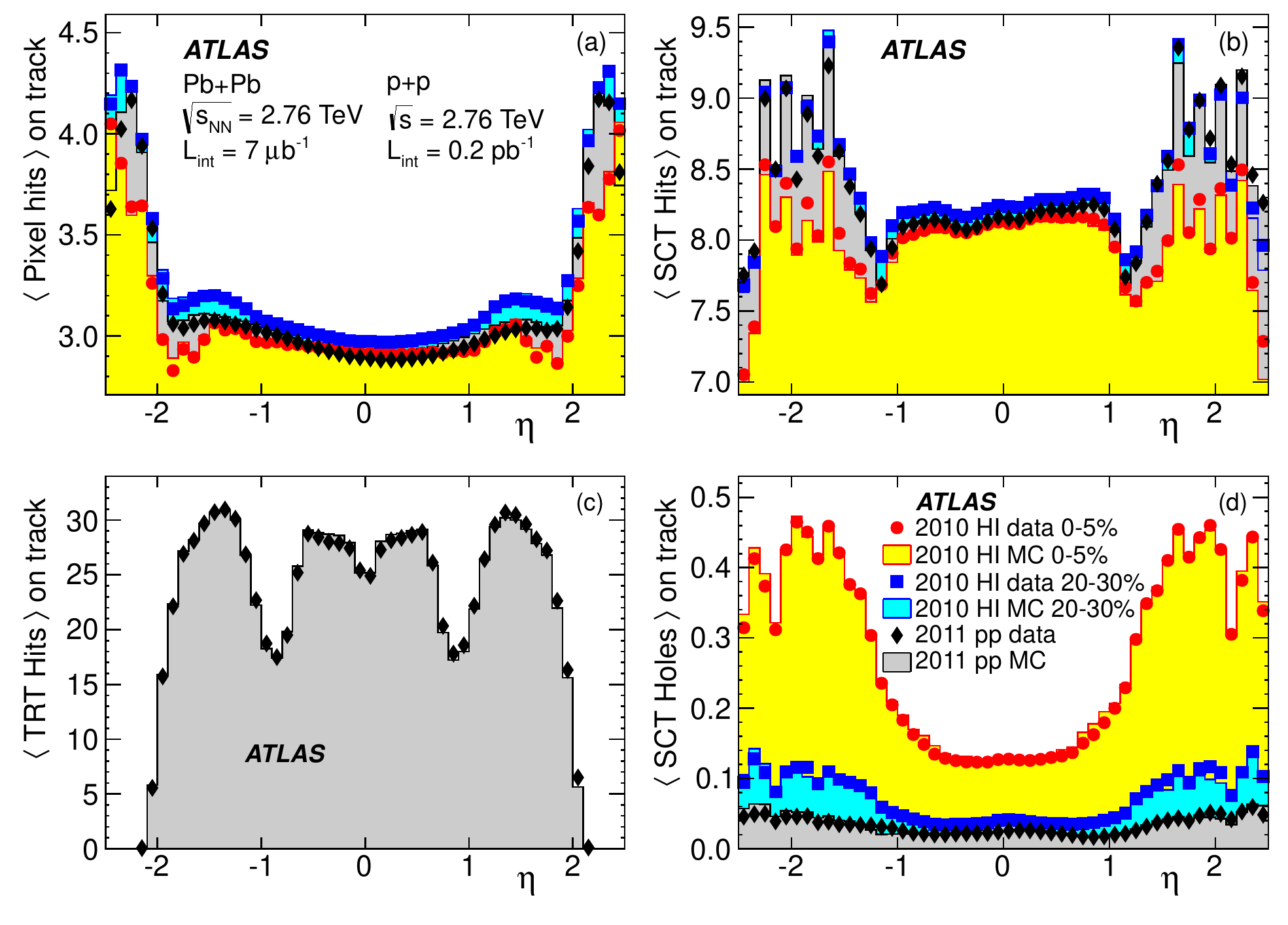}
	\caption{(a) The average number of Pixel hits per track, (b) the average number of SCT hits per track, (c) the average number of TRT hits per track, (d) the average number of SCT holes per track, measured in \PbPb events in 0--5\% (circles) and 20--30\% (squares) centrality classes and in \pp\ events (diamonds) as a function of $\eta$. The data from the 2010 \PbPb\ run and 2011 \pp\ runs are compared to the corresponding distributions obtained from simulations shown with filled histograms.}
\label{fig:on_track}
\end{figure*}

The simulation reasonably reproduces the centrality dependence of the track reconstruction parameters which define the quality of the track. Figure~\ref{fig:on_track} shows the average number of Pixel, SCT and TRT hits and the average number of SCT holes in \PbPb events for the 0--5\% and 20--30\% centrality classes and in the \pp\ data. An SCT hole is defined by the absence of a hit, predicted by the track trajectory, in a given detector layer. The prediction takes both the detector geometry and the active detector area into account. In HI data processing, tracks with holes in the Pixel detector are rejected at earlier stages by the ID track reconstruction. The number of TRT hits is shown only for the \pp\ data, where it plays the most important role. 

All parameters shown in figure~\ref{fig:on_track} are sensitive to the momentum distribution of the tracks; therefore the MC tracks were reweighted in \pT\ to match the distributions measured in the data. The TRT covers the pseudorapidity interval $|\eta|<2$, which limits the analysis of the \pp\ data and \rpbpb to this interval, but the TRT hits requirement is necessary to improves track sample purity at high \pT. To avoid possible bias at low \pT, only tracks above \pT\ of 6\,\GeV\ are required to have hits in the TRT. The results with and without such requirement are consistent in the 2--10\,\GeV\ interval. 

The mean number of Pixel and SCT hits as a function of $\eta$ shows a relatively weak centrality dependence, whereas the number of SCT holes increases significantly with centrality. This trend shows the effect of the large occupancy on the SCT pattern-recognition algorithm. A hole on a track may appear if a hit in the SCT is assigned to another track. The fake-track rejection in \PbPb collisions requires that there is no hit sharing between tracks. An increased number of SCT holes is an indication of a deterioration of the track quality. Tracks where a SCT hole is present are rejected from the analysis to limit the fake rate. 
 
Tracks are selected for the analysis if they have $\pT>0.5$\,\GeV\ and satisfy all of the quality requirements summarized in table~\ref{table:trk_req}. In addition, tracks are required to have $|\eta|<2.5$ in \PbPb samples and $|\eta|<2.0$ in \pp samples.

\begin{table*}[tb]
\begin{center}
\begin{tabular}{|c|c|cc|} \hline
 	\multicolumn{4}{|c|}{\PbPb requirements} \\ \hline
		& Basic & \multicolumn{2}{c|}{For systematic uncertainty} \\ \hline
	Pixel hits & $\geq 2$, BL hit & $\geq 1$, BL hit (if exp.) & $\geq$ 3 layers \\
	SCT hits & $\geq 7$  & $\geq 6$ & $\geq 8$  \\
	SCT holes & =0 &  & \\  \hline
 \end{tabular} 
 \vspace*{0.3cm}\\
 \begin{tabular}{|c|c|cc|}  \hline
 \multicolumn{4}{|c|}{\pp requirements} \\ \hline
			& Basic & \multicolumn{2}{c|}{For systematic uncertainty} \\ \hline
		Pixel hits & $\geq 1$, BL hit (if exp.) & & $\geq 2$\\
		SCT hits & $\geq 6$  & $\geq 5$ & $\geq 7$  \\
		SCT holes & $\leq$1 & \hspace*{1.95cm} & \hspace*{1.95cm}\\ 
		TRT hits & $\geq 8$ & & $\geq10$ \\ \hline
 \end{tabular}
\caption{Track selection criteria for \PbPb and \pp events. ``Basic" cuts are used by default in the analysis. Other cuts are used only for estimation of systematic uncertainties. ``BL hit" denotes for the requirement of a hit in the innermost pixel layer, so-called B-layer. ``BL hit (if exp.)" denotes for the same requirement, but only if such a hit is expected by the track reconstruction algorithm. The TRT hit requirement is applied only to tracks with $\pT>6$\,\GeV.}
\label{table:trk_req}
\end{center}
\end{table*}

\subsection{Requiring pointing to the primary vertex}
\label{sec:pv_match}
In \PbPb collisions the vertex finding and fitting algorithm~\cite{pp_pv_perf} provides accurate primary vertex (PV) reconstruction. In the most peripheral collisions used in this analysis, the PV reconstruction precision is better than 20\,$\mu$m in the transverse plane, allowing for efficient separation of tracks that are produced in primary interactions from those that arise from later decays. In more central collisions the PV precision improves inversely proportional to the square root of the number of reconstructed primary tracks in the events. Since the rate of in-time pile-up events in the 0--5\% \PbPb collisions was less than $5 \times 10^{-3}$, only one PV is considered in each event. For each track, the distance of closest approach to the PV is determined in the transverse and longitudinal directions (\dz\ and \zz\ respectively), along with the errors of these quantities (\edz\ and \ezz) obtained from the covariance matrix of the track fit. They allow for efficient discrimination of tracks arising from secondary particles or tracks with poorly measured \pT.

\begin{figure*}[!b]
\centering
\includegraphics[width=0.95\textwidth]{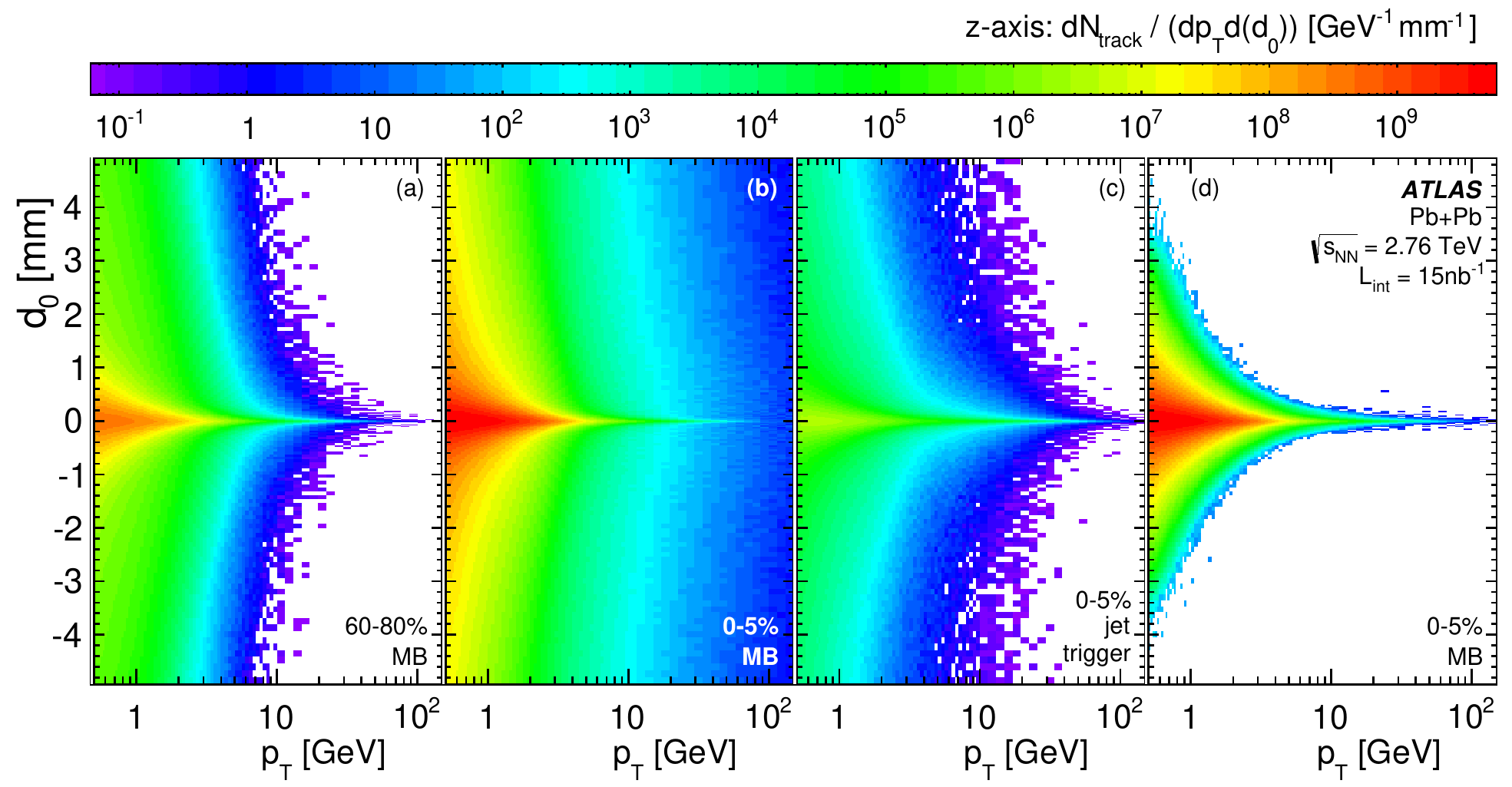}
\caption{Distance of closest approach of the track trajectory to the PV in the plane transverse to the beam direction, \dz, (a) in 60--80\% peripheral collisions of the MB sample, (b) in 0--5\% central collisions of the MB sample, (c) in 0--5\% central collisions for the tracks matched to jets in the jet-triggered sample, (d) in \mbox{0--5\%} central collisions of the MB sample after applying cuts on the \dz\ and \zz\ significances (see text). The \mbox{$z$-axis} $\mathrm{d}N_\mathrm{track}/(\mathrm{d}\pt \mathrm{d}\dz)$ uses a logarithmic scale and is the same for all panels. White areas in the distributions are at least eleven orders of magnitude below the maximum in panel (b).}
\label{fig:pv_pointing}
\end{figure*}

Figure~\ref{fig:pv_pointing} shows scatter plots of the \dz\ parameter versus \pT\ under different conditions. The distributions of \zz, not shown here, are similar. The widths of the \dz\ distributions increase with decreasing \pT\ due to the increase of multiple scattering.  All tracks satisfy the selection criteria listed in section~\ref{sec:track_params}. Figure~\ref{fig:pv_pointing}(a) shows the \dz\ versus \pT\ scatter plot for all tracks in the MB events in the 60--80\% centrality interval. The width of the $\dz$ distribution decreases with increasing transverse momentum of the tracks. At very high \pT\ it approaches the limit of about 10\,$\mu$m defined by the Pixel detector resolution. At $\pt<2$\,\GeV\ and large $|\dz|$ a significant fraction of tracks originate from weak decays (mainly $\mathrm{K}^0_\mathrm{S}$ and $\Lambda^0$) inside the pixel detector. At higher \pT, the region with $|\dz|$ exceeding 100\,$\mu$m is unpopulated. The distribution of track parameters shown in figure~\ref{fig:pv_pointing}(a) is similar to that in \pp collisions, since both systems have relatively low occupancy and thus similar ID tracking performance.

The \dz\ distributions in the 0--5\% most central collisions, shown in figure~\ref{fig:pv_pointing}(b), are different from those in peripheral collisions.  In central collisions, the \dz\ distribution extends to large values even for high-\pT\ tracks. To understand the origin of these tracks it is necessary to compare figure~\ref{fig:pv_pointing}(b) to a distribution obtained from a sample of events measured under the same occupancy conditions obtained for the jet-triggered sample, shown in figure~\ref{fig:pv_pointing}(c).

The tracks included in figure~\ref{fig:pv_pointing}(c) satisfy the condition $\Delta R=\sqrt{\left(\Delta\eta \right)^{2}+\left(\Delta\phi \right)^{2}}<0.4$, where $\Delta\eta$ and $\Delta\phi$ are the difference in pseudorapidity and azimuthal angle respectively between the directions of the tracks and the jet axis. Matching tracks to jets is further described in section~\ref{sec:jet_match}. Since most tracks with low \pT\ are produced via soft interactions, the yield of these tracks is suppressed by the angular matching to jets.

At high \pT, the tracks matched to jets represent a sample with relatively fewer fake tracks compared to the one shown in figure~\ref{fig:pv_pointing}(b), which is characterized by the presence of large \dz\ values at high \pT. The shape of the \dz\ distribution is similar to that seen in peripheral collisions. This leads to the conclusion that in central collisions, tracks populating the region at high \pT\ and large $|\dz|$ in figure~\ref{fig:pv_pointing}(b) are not due to a modification of the \dz\ distribution of high-\pT\ tracks, but rather to mismeasured momentum of low-\pT\ charged particles caused by increased occupancy in the SCT detector. The \dz\ and \zz\ parameters of the tracks remain properly measured by the Pixel detector, which has a low occupancy even in most central \PbPb collisions. 

\begin{figure}[!t]
\centering
\includegraphics[width=0.7\columnwidth]{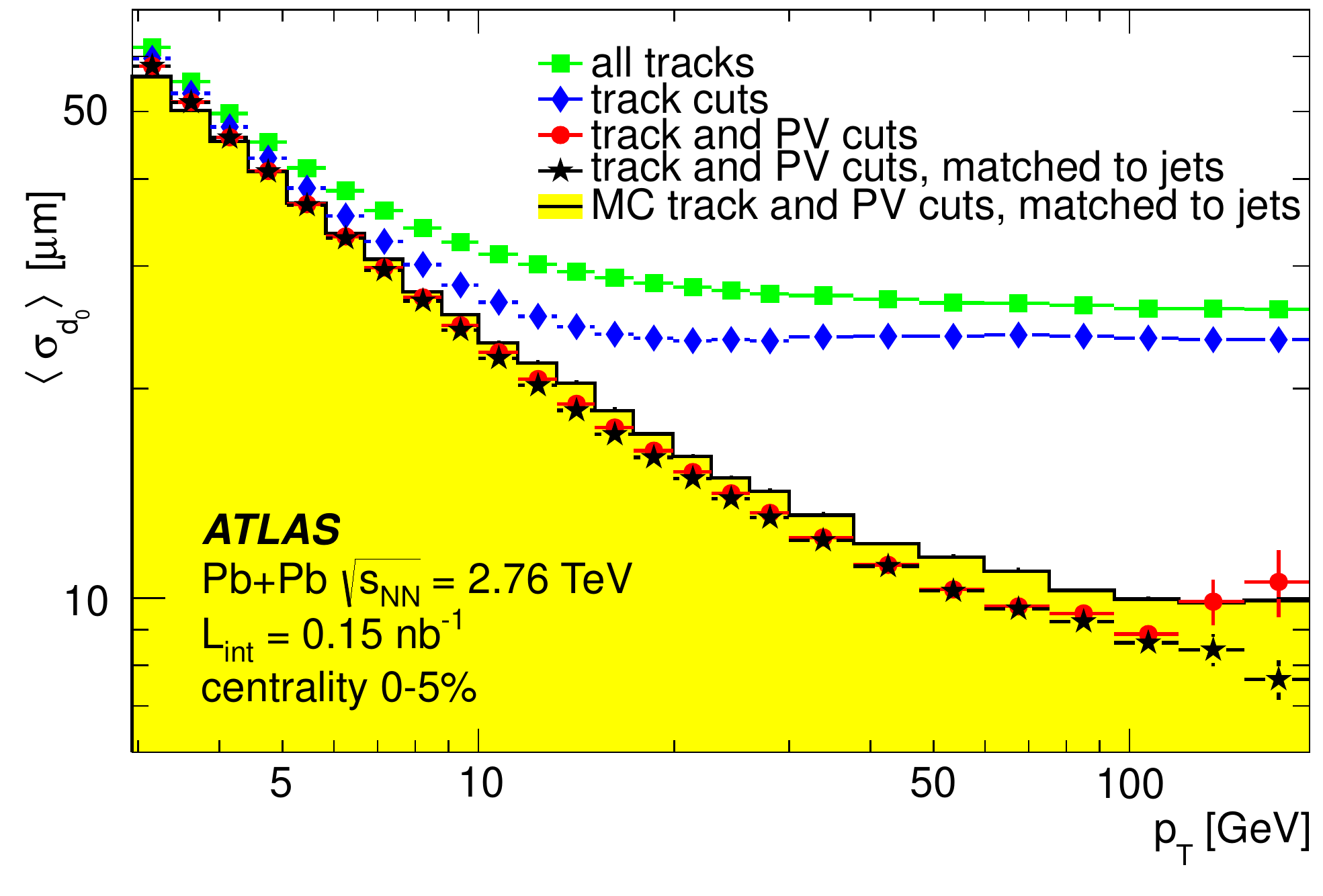}
\caption{The average uncertainty of the transverse impact parameter, $\langle\edz\rangle$, as a function of track \pT, measured in the 0--5\% central collisions. Squares represent all tracks before applying cuts. Diamonds denote tracks after track selection requirements (section~\ref{sec:track_params}). The circles, mostly overlapping the stars, are for the tracks satisfying PV pointing requirements, and stars are the same for the tracks also matched to jets above $\pt>20$\,\GeV\ (explained in section~\ref{sec:jet_match}). The filled histogram is the result for the simulated tracks from primary particles plotted as a function of reconstructed track \pT\ and satisfying the same requirements as in the data.}
\label{fig:ed0}
\end{figure}

Accurate estimates of the impact parameter resolution, \edz\ and \ezz, use all hits, including the SCT hits which define the momentum of the track. An overestimated momentum of the track corresponds to a straighter trajectory with lower multiple scattering. The \edz\ and \ezz\ for such mismeasured tracks are overestimated with respect to the measured \pT.  This effect is seen in figure~\ref{fig:ed0}, which shows the average value of the $\dz$ uncertainty, $\langle\edz\rangle$, as a function of track $\pT$, measured in the 0--5\% most central collisions. Averaging is done over all tracks in the MB sample. The squares represent all reconstructed tracks. Above $\pT$ of 10\,\GeV\ the $\langle\edz\rangle$ distribution changes very little with \pT\ due to contamination by tracks from low-\pT\ charged particles with badly measured \pT. Applying the track selection requirements explained in section~\ref{sec:track_params} reduces the average \edz\ as shown by diamond symbols (the same sample as shown in figure~\ref{fig:pv_pointing}(b)). At high \pT, $\langle\edz\rangle$ is still much larger than that for tracks in the jet-triggered sample which are shown with stars.

From these considerations, it can be concluded that the significances, $\dz/\edz$ and $\zz/\ezz$, for mismeasured tracks from low-\pT\ particles are much larger than for correctly reconstructed tracks. Rejecting tracks with absolute value of either the transverse or longitudinal impact parameter significance higher than 3.0 from the sample of tracks shown in figure~\ref{fig:pv_pointing}(b) produces the scatter-plot presented in figure~\ref{fig:pv_pointing}(d). The distribution at high \pT\ more closely resembles the distribution shown in figure~\ref{fig:pv_pointing}(a), except for the low-\pT\ tracks, where the cuts on the impact parameter significance reject a large fraction of secondary particles. The $\langle\edz\rangle$ for these tracks, shown with circles in figure~\ref{fig:ed0}, is much closer to the values for tracks in jets up to track \pT\ of 100\,\GeV\ and to the results of the MC simulation shown in the figure as a histogram.

Tracks are selected for analysis in \PbPb samples if they satisfy the requirements \mbox{$|\dz|/\edz<3.0$ and $|\zz|/\ezz <3.0$.} Otherwise tracks are not considered to originate in the primary vertex. In addition $\edz/\langle\edz\rangle$ is also required to be $<3.0$. 

In \pp\ collisions the PV precision in the transverse plane is about 80\,$\mu$m for events with 20 tracks and improves to 30\,$\mu$m for events with 70 tracks~\cite{pp_pv_perf}. The presence of a jet in an event typically increases the vertex reconstruction precision by contributing tracks. However, using PV pointing cuts similar to those used in the \PbPb analysis introduces a correlation between the track selection criteria and the jet multiplicity, which can bias track selection. Therefore, only a loose requirement of $|\dz|<1.5$\,mm is used in the \pp\ track selection to suppress the contribution of secondary tracks.

\subsection{Matching tracks to calorimetric jets}
\label{sec:jet_match}
Tracks are required to be matched to jets in order to remove the contribution of fake tracks at high \pT. Calorimetric jets are reconstructed using the \akt algorithm with the radius parameters $R=0.2, 0.3, 0.4$~and~0.5. In HI collisions, the jet kinematics are obtained by performing the reconstruction on $\Delta\eta \times \Delta\phi = 0.1 \times 0.1$ towers prior to underlying-event subtraction and then iteratively evaluating and subtracting the estimated underlying-event contribution from each jet at the calorimeter cell level, accounting for the effects of elliptic flow modulation in \PbPb collisions~\cite{ell_flow}. The details of the jet reconstruction in \PbPb collisions are described in ref.~\cite{jet_paper_new}. The inputs to the jet algorithm are topological clusters of calorimeter cells without a background subtraction procedure, as described in ref.~\cite{jet_atlas_pp}. In both the HI and \pp\ collisions, jets in jet-triggered samples are required to be matched to the online reconstructed jets used by the corresponding jet trigger. This ensures that the tracks correspond to the jets which fired the trigger. No such requirement was used for jets in the MB samples.

\begin{figure}[!tb]
\centering
\includegraphics[width=0.7\columnwidth]{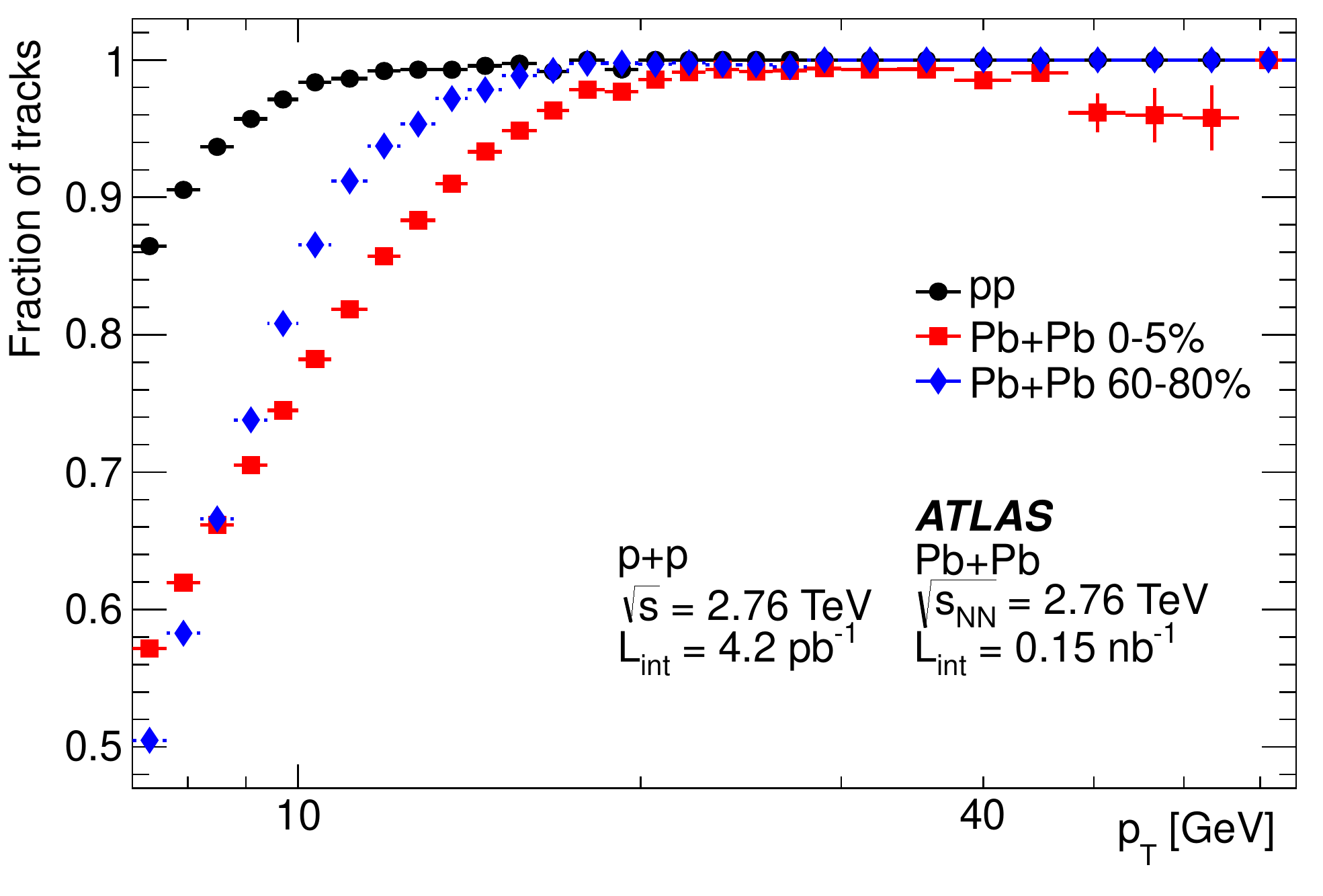}
\caption{The fraction of tracks in the MB samples, within all tracks in such sample, residing in a cone of size $\Delta R=0.4$ around the direction of a jet as a function of the track \pT. All tracks satisfy the requirements listed in section~\ref{sec:track_params} and \ref{sec:pv_match}. Circles represent \pp\ data using $R=0.4$ jets. Squares and diamonds represent \PbPb data using $R=0.2$ jets in the centrality intervals 0--5\% and 60--80\%, respectively.}
\label{fig:matching}
\end{figure}

High-\pT\ tracks are produced by particles formed during parton fragmentation to jets, except for those produced by leptons ($\ell$) coming from decays of electroweak bosons, primarily $W^{\pm}\rightarrow \ell^{\pm}\nu$ and $Z\rightarrow \ell^{+}\ell^{-}$. The contribution of leptons from $W$ and $Z$ decays is removed from the data by identifying tracks corresponding to muons in the ATLAS muon spectrometer and assuming the contribution of electrons is the same. The contribution of leptons is the largest at $\pT$ around 40\,\GeV\ and in the most central \PbPb collisions where it reaches 5\%. The matching of tracks to jets is studied in the data. Figure~\ref{fig:matching} shows the fraction of tracks, selected by applying the requirements described above, which reside in a cone of size $\Delta R=0.4$ centred on the jet axis in the \PbPb and the \pp\ MB samples. At $\pT<10$\,\GeV\ many tracks cannot be matched to jets due to the limited centrality-dependent reconstruction efficiency for jets with low $\pT$ \cite{jet_triggers,jet_xs_pp}. At $\pT \gtrsim 15$\,\GeV\ all tracks reconstructed in the \pp\ data, shown with circles, reside within the cone around the direction of a jet. In peripheral \PbPb collisions, shown with diamond symbols, all tracks with $\pT \gtrsim 20$\,\GeV\ are within the cone of a jet. In central collisions, shown with squares, in the same \pT\ range about 99\% of all tracks are found within the cone of a jet; however, above 40\,\GeV\ this fraction diminishes due to the contributions of misreconstructed tracks. 

\begin{figure*}[!tbh]
\centering
\includegraphics[width=0.7\textwidth]{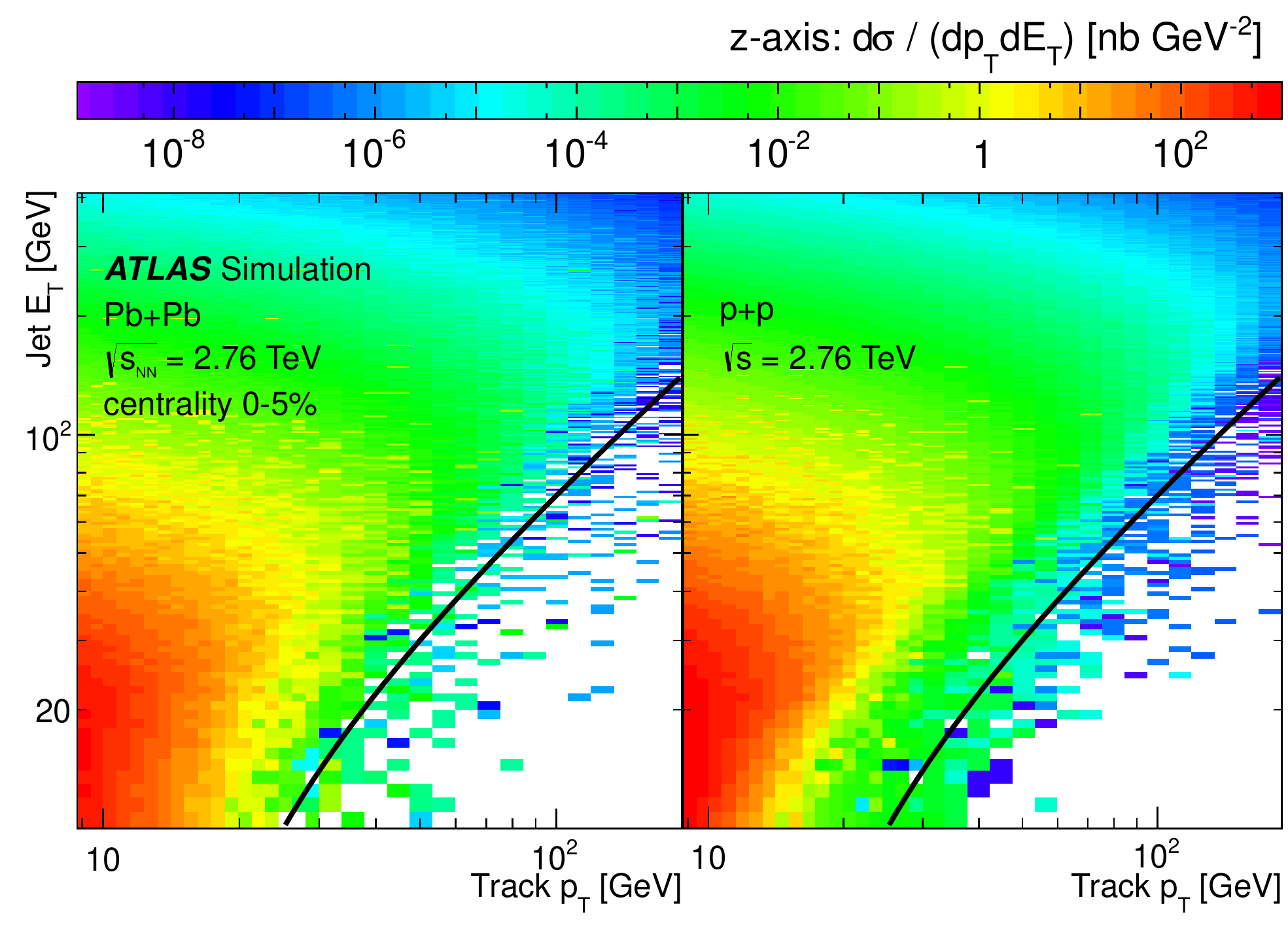}
\caption{The yield of tracks with reconstructed momentum \pT\ matched to a jet with measured transverse energy \et. The left panel shows the \PbPb yield divided by $\avgTaa$ obtained by embedding PYTHIA events into the 0--5\% most central \PbPb\ data. The right panel shows the \pp cross section as simulated by PYTHIA. Track and jet selections are explained in the text. Lines correspond to the expression $\et=0.8\times\pT-10$\,\GeV.}
\label{fig:track_jet}
\end{figure*}

Tracks with $\pT>15$\,\GeV\ in \pp\ collisions and $\pT>20$\,\GeV\ in \PbPb collisions both in the MB and jet-triggered samples are required to be within a cone of $\Delta R=0.4$ around the jet direction. The jet is reconstructed with a radius parameter of $R=0.2$ in \PbPb and $R=0.4$ in \pp\ collisions.
 
In addition to the requirement of an angular matching, the track \pT\ is also compared to the \et\ of the jet. The track-to-jet energy balance was studied in the MC simulation. The results are sensitive to the description of parton fragmentation in simulation. In \PbPb\ data the jet fragmentation is modified due to medium effects~\cite{frag_atlas} that are not included in the PYTHIA generator. Therefore, the track-to-jet energy balance cut is chosen such that it only rejects very unbalanced pairs of tracks and jets. 

The results are shown in figure~\ref{fig:track_jet} as a two-dimensional distribution of the track \pT\ versus \et\ of the jet in the 0--5\% most central \PbPb\ collisions on the left and in \pp\ collisions on the right. The histograms show the \et\ of the jet to which the track is matched vs.~the track \pT. The reconstructed \pT\ of the track is typically lower than the measured jet energy, reflecting the fragmentation process. Matches of tracks to jets with \et\ less than the track \pT\ result from the finite resolution of the jet energy measurement or track momentum reconstruction, or from badly reconstructed or fake tracks. The solid line shown in the plot corresponds to $\et=0.8\times\pT-10$\,\GeV\ which was chosen as the cut of the track-to-jet energy balance.

\subsection{Merging triggered data samples}
\label{sec:merge}

The samples obtained with different triggers are combined to construct the spectra, which are equivalent to the unbiased MB spectra and extended to the highest \pT\ values allowed by the luminosity. This is done separately for each data-taking period. The simulation samples are merged by summing the hard scattering and MB distributions according to their cross sections. The data distributions are corrected using the summed MC samples corresponding to the given run period. After applying corrections described in section~\ref{sec:correction}, the data samples are merged from different running periods.

In the data, tracks with \pT\ low enough to not require matching to a jet as explained in section~\ref{sec:jet_match}, are taken entirely from the MB samples. They have the proper mixture of all processes presented in the collisions. 

Tracks with \pT\ above the track-to-jet matching requirement are taken from both the MB and jet-triggered samples. These tracks are required to be matched to a jet, including those from MB samples. Jet-triggered samples are used when the corresponding jet trigger efficiency reaches 90\%. Efficiencies of the jet triggers defined with respect to the MB triggers are evaluated in refs.~\cite{jet_triggers,jet_xs_pp}. The jet trigger prescale factor is smaller for triggers with higher thresholds. When a jet trigger with a higher threshold also reaches 90\% efficiency its sample is larger than the sample with lower jet threshold, and therefore the use of the jet sample with the lower threshold trigger is discontinued. This procedure guarantees that each track matched to a jet is used only once in all samples, even if the same jet was recorded by more than one trigger. 

\begin{figure}[!b]
\centering
\includegraphics[width=0.7\columnwidth]{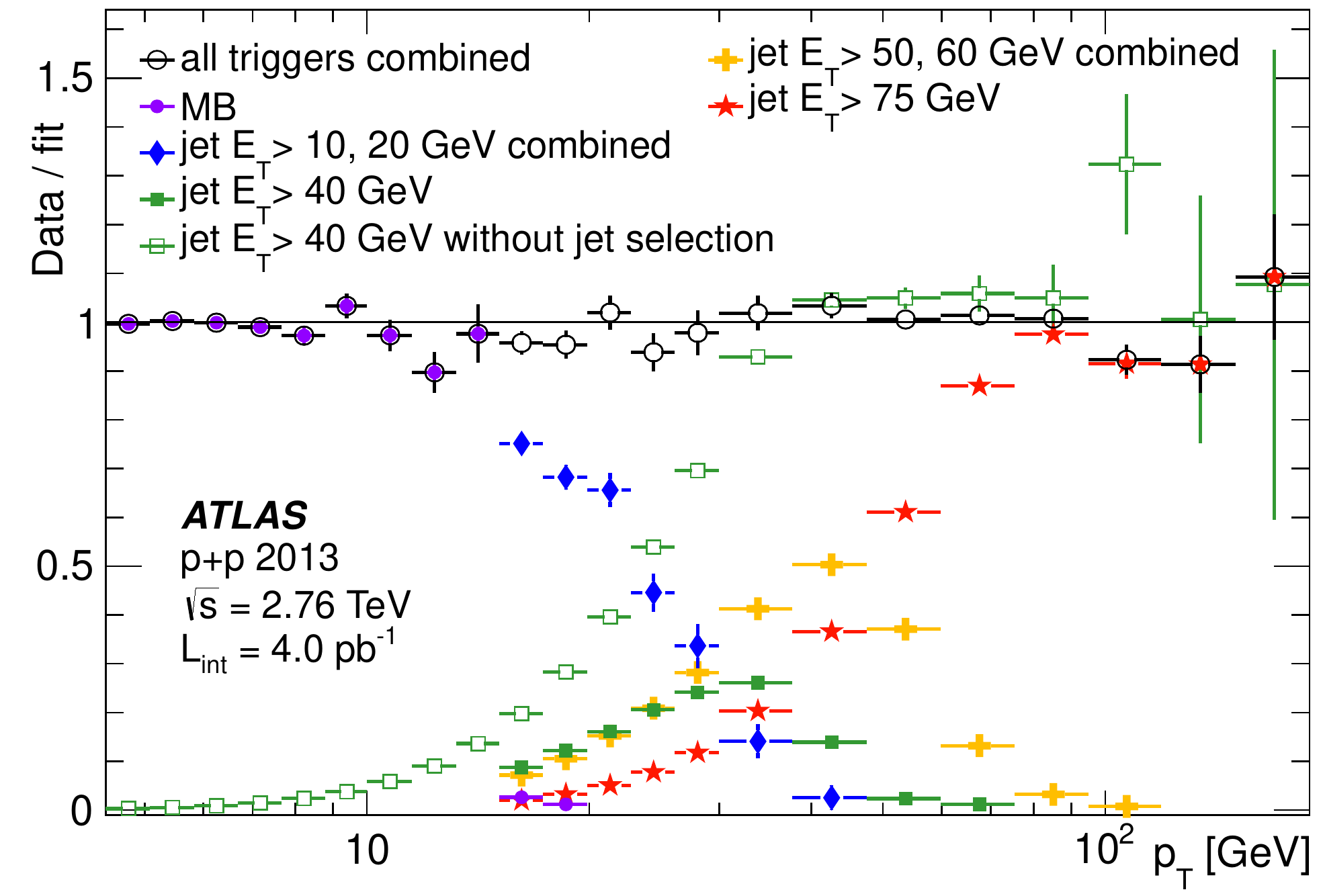}
\caption{Ratios of the number of tracks from different \pp\ triggered samples to a smooth parameterization (``fit") describing the combined sample. Open circles show the ratio for the combined sample of all tracks. Filled circles show the MB sample. Other filled markers show the contributions of tracks in jet-triggered samples. Only one sample is used for all jets with the same \ET\ (see text). Some samples are combined for visibility of the plot. Open squares show the total number of tracks in the sample of the jet trigger $\et>40$\,\GeV.}
\label{fig:trig_samples}
\end{figure}

After correcting for average prescale of each sample, the selected tracks are added to produce a single combined sample as explained above. The contributions of triggered samples to the combined sample are shown in figure~\ref{fig:trig_samples} for the 2013 \pp\ data. The data are divided by a power-law parameterization to better reveal the contributions from each sample. Open circles around unity depict the combined sample of all tracks. The contribution of the MB sample below the track-to-jet matching \pT\ threshold (15\,\GeV), shown with filled circles, coincides with the combined distribution. At higher \pT, the tracks from the MB sample contribute only when they are matched to low-energy jets. This part is also shown with filled circles.

At even higher \pT, tracks come predominantly from the event samples obtained with jet triggers. Their \pT\ distributions are shown with filled markers; some trigger samples are combined for plot clarity. The contributions from such samples first rise and then decrease, except for the two lowest-threshold trigger samples where only the decrease is presented, and for the highest-threshold trigger sample with $\et>75$\,\GeV. Since jets are selected from each trigger only in an exclusive \et\ range, the contribution of tracks to the total spectrum rises and falls with \et. The effect of such a jet trigger interval selection is demonstrated by comparison of closed squares (with selection) and open squares (without selection) corresponding to the jet trigger with $\ET$ threshold of 40\,\GeV. Without applying a jet trigger interval selection the ratio starts at low \pT\ at a value significantly below unity. The ratio then rises to unity where it remains flat, consistent with the combined sample of all tracks. However, the statistical uncertainty of the ratio is significantly larger than that for the samples obtained with the higher jet trigger thresholds due to the difference in the integrated luminosity associated with each trigger.

\section{Acceptance and efficiency corrections} 
\label{sec:correction}
Corrections to the reconstructed yields in each sample are applied at different stages of the analysis in order to remove detector effects, reconstruction biases and to correct to the particle level. They are derived from data and simulation. 

The correction for the fraction of events recorded with different jet and MB triggers is assigned as a weight to each track selected for the analysis. The jet trigger efficiency correction is applied in the same way. The trigger efficiencies are measured in analyses published in refs.~\cite{jet_triggers, jet_xs_pp}. This correction is relatively small, since each jet trigger is used only to select jets for which it is at least 90\% efficient. The correction for the MB trigger efficiency in \PbPb collisions is negligible in all centrality intervals analysed. In the \pp\ data a correction for the vertex reconstruction efficiency is applied. The vertex reconstruction efficiencies are estimated as a function of number of reconstructed tracks in the same way as in ref.~\cite{ATLASminb}. The correction is very small and reaches unity when four or more tracks are selected within an event. After applying these corrections, samples are merged as described in section~\ref{sec:merge}.

\begin{figure}[!tb]
	\centering
	\includegraphics[width=0.7\columnwidth]{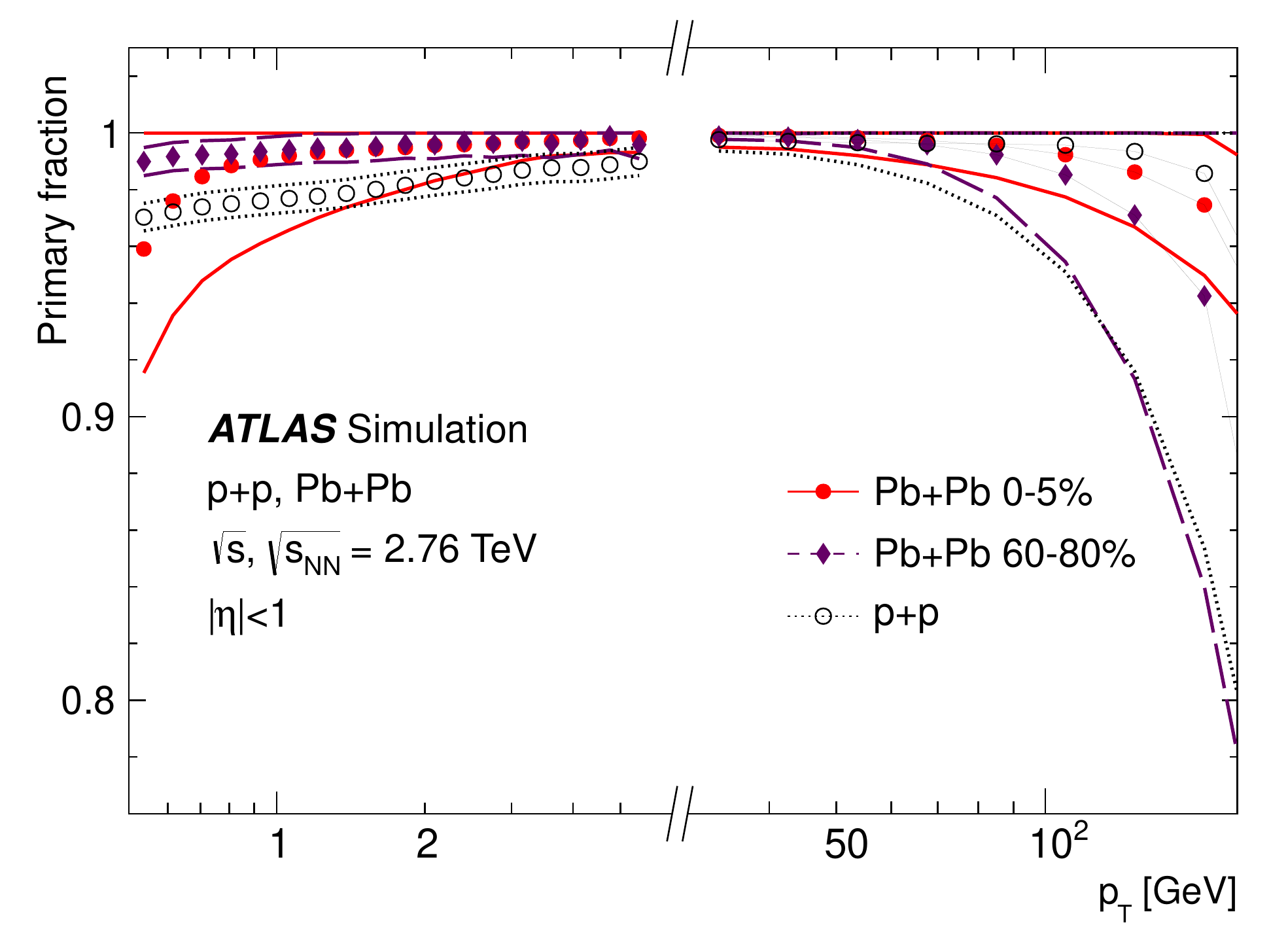}
	\caption{The fraction of reconstructed tracks matched to primary particles for two \PbPb\ centrality classes (solid markers) and \pp\ (open markers) as a function of track \pT. Their systematic uncertainty bands are shown with solid, dashed and dotted lines for \PbPb central collisions, peripheral collisions and \pp, respectively, and are further discussed in section~\ref{sec:syst_corr}.}
	\label{fig:prim_frac}
\end{figure}

\begin{figure}[!tb]
	\centering
	\includegraphics[width=0.7\columnwidth]{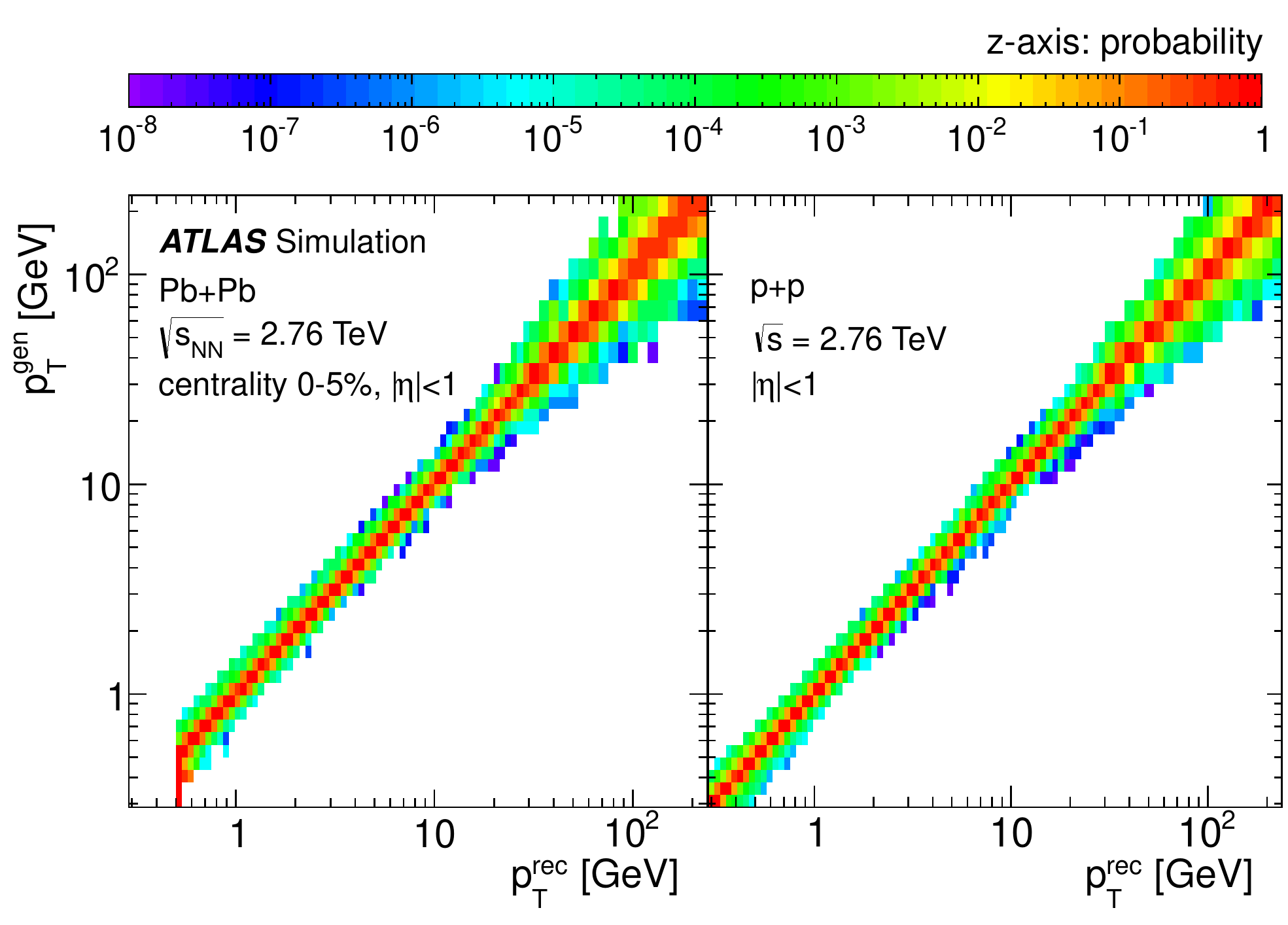}
	\caption{Examples of migrations from generated $\pT$ ($\pT^\mathrm{gen}$) to reconstructed $\pT$ ($\pT^\mathrm{rec}$) for tracks in the most central collisions in \PbPb\ (left) and for \pp (right). The integral of the distribution in each $\pT^\mathrm{gen}$ bin is normalized to unity.}
	\label{fig:mig_matr}
\end{figure}

After the raw spectra from different triggers are merged, they are corrected for secondary, fake and misreconstructed tracks. The merged spectra are multiplied by the fraction of primary tracks, which is defined as the number of primary tracks divided by the total number of tracks, i.e. including secondary, fake and misreconstructed tracks. The fractions are estimated in the MC simulation for 20 bins in $\eta$ with a width of 0.25, 9 centrality intervals, and separately for positively and negatively charged particles. They are found to have only a weak dependence on particle charge and the sign of $\eta$. However, they are dependent on $|\eta|$ for $|\eta|>1$. Thus, the corrections were merged into seven $|\eta|$ bins for both charges combined in order to have smaller statistical uncertainty at higher \pT. At low \pT\ this correction is dominated by secondary particles and reaches 10\% at $|\eta|>2.25$ in the most central \PbPb collisions. It decreases at low $|\eta|$ and in more peripheral collisions. In the \pp\ data samples it does not exceed 4\% at any measured $\eta$. At $\pT>4$\,\GeV\ the contributions of fakes and secondaries do not exceed 1\% in any of the samples and remain at this value up to $\pT \approx 70$\,\GeV. At the highest measured \pT, misreconstructed tracks contribute significantly, as shown in figure~\ref{fig:prim_frac}.

The spectra are then corrected for the track momentum resolution with an iterative Bayesian unfolding procedure \cite{BayesUnf}. Results are obtained with two iterations. The migration matrices used are shown in figure~\ref{fig:mig_matr} as functions of the generated particle transverse momentum ($p_{\mathrm{T}}^\mathrm{gen}$) and reconstructed transverse momentum of the associated track ($p_{\mathrm{T}}^\mathrm{rec}$). The migration matrices are consistent for both particle charges and for the same $\eta$ ranges as for the primary fraction correction and are therefore combined.

\begin{figure}[!tb]
	\centering
	\includegraphics[width=0.7\columnwidth]{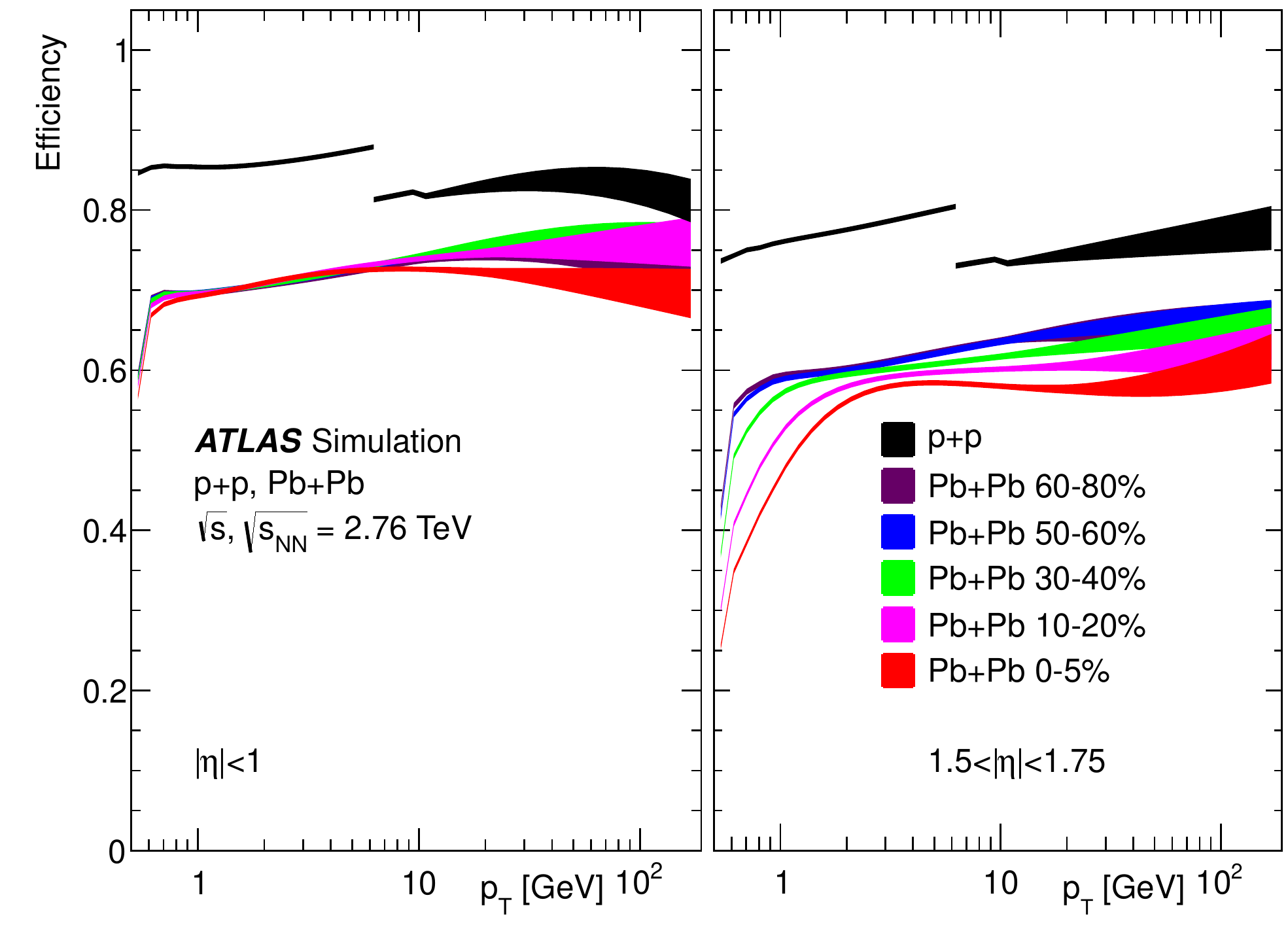}
	\caption{Track reconstruction efficiency as a function of $\pT$ for five centrality classes for \PbPb\ and for \pp in two $|\eta|$ regions. The discontinuity of the \pp\ efficiency at 6\,\GeV\ is caused by the change in the TRT hit requirements (see section~\ref{sec:track_params}). The width of the band represents the systematic uncertainty.}
	\label{fig:efficiency}
\end{figure}

The unfolded spectra are corrected for the reconstruction efficiency using the efficiency estimated in simulation, defined as the number of primary charged particles matched to reconstructed tracks divided by the total number of generated primary particles. The track reconstruction efficiencies are estimated as a function of reconstructed particle \pT, $|\eta|$ and charge. They are found to agree in the same regions as the primary fractions and are also combined for both charges of particles, both sides of $|\eta|$ and for $|\eta|<1$. The efficiencies are smoothed by fitting to minimize the effect of statistical fluctuations at high \pT\ in the simulation. 
A sample of efficiency curves is shown in figure~\ref{fig:efficiency}.

\section{Systematic uncertainties}
\label{sec:syst}
The systematic uncertainties of the measurements presented in this \papertype arise from several sources, which are summarized at the end of this section in table \ref{table:sysErr}. Uncertainties related to the absolute normalization of the spectra come from the uncertainties of the integrated luminosity of the \pp\ sample and event selection efficiency in \PbPb. The systematic uncertainties related to the event selection include the uncertainties on the jet trigger efficiencies used in the \PbPb and \pp\ event samples. Uncertainties related to the selection of tracks include possible differences in the data and in the MC simulation. Finally, there are uncertainties in the determination of the applied corrections.

Systematic uncertainties are evaluated by varying parameters in the analysis for each source and observing the effects on the corrections and yields. Variations are applied simultaneously to the \PbPb and \pp data, so that final uncertainties on \raa and \rcp properly reflect the correlated changes in the numerator and denominator. The total systematic uncertainties are determined by adding the uncertainties from each source in quadrature.

\subsection{Absolute normalization}
The uncertainty on the integrated luminosity of the \pp\ samples is 3\% and is taken into account in the invariant cross section and in the \raa measurements presented in this \papertype. It was estimated in a fashion similar to that previously described by ATLAS~\cite{lumi_pp7}. The uncertainty on the fraction of events in the MB sample in \PbPb collisions is evaluated in ref.~\cite{multiplicity} to be 2\% for both the 2010 and 2011 runs. This uncertainty affects both the boundaries of the centrality intervals and the centrality-related geometric parameters, $\avgTaa$. The largest effect is in the 90--100\% centrality interval, which is not considered in the paper. To avoid double counting in the results presented in this \papertype, the uncertainty due to the event fraction in the MB sample is included only in the uncertainties of $\avgTaa$ and $\avgNpart$ as given in table~\ref{table:centrality}.

\subsection{Event selection}
The uncertainties on the determination of the jet trigger efficiencies in different jet samples in the \pp\ sample are evaluated in the analysis published in ref.~\cite{jet_xs_pp}. The resulting uncertainty on the spectra from this contribution does not exceed 3\%. 

In 2011 the \PbPb events were recorded if the \akt jet reconstruction algorithm found a jet with radius parameter $R=0.2$ and $\eT>20$\,\GeV\ after underlying-event subtraction. The trigger efficiency in \PbPb events was studied in the inclusive jet analysis published in ref.~\cite{jet_triggers}. This reaches 90\% for jets with $\eT>40$\,\GeV\ and has a weak centrality dependence. The systematic uncertainties on the charged-particle spectra related to the jet trigger efficiency uncertainty is 1\%, less than in \pp, because only one jet trigger was used in the \PbPb data-taking.

\subsection{Track selection}
The uncertainty due to the track selection reflects possible differences in performance of the track reconstruction algorithms in data and in MC simulation. To estimate this uncertainty, the track quality requirements described in section~\ref{sec:track_params} and the vertex pointing requirement described in section~\ref{sec:pv_match} were tightened or loosened in both data and MC simulation. The resulting uncertainty is below 1\% at \pT\ below 10\,\GeV\ and rises with increasing \pT, reaching a maximum of 10\% at the highest measured \pT\ for \PbPb and 4\% for \pp collisions.

Fake and secondary particles passing the tracking requirements create a small contamination at low \pT\ and are accounted for through the primary fraction correction. The uncertainty of this contamination is due to the unknown relative contributions of primary, secondary and fake tracks. The uncertainty is estimated by comparing the fraction of primary tracks, as defined in this \papertype, to the fraction of tracks with a reconstructed impact parameter consistent with the particle arising from the primary vertex. In central collisions, the uncertainty is 5\% at the lowest measured \pT\ and decreased to 0.5\% at 1\,\GeV, in all centralities.

The uncertainty due to fake and secondary particles production at high \pT\ is highly correlated with the uncertainty associated with the procedure for matching generator-level particles to reconstructed tracks. The uncertainty is at most 3\% at the highest measured \pT. In table~\ref{table:sysErr}, the high-\pT\ part of the uncertainty on fake and secondary particles is included in the uncertainty of the matching.

\subsection{Correction procedure}
\label{sec:syst_corr}
These uncertainties reflect the possible impact of variations in the derivation of the applied corrections. The uncertainty associated with matching generated particles to reconstructed tracks is correlated with the uncertainty associated with fake and secondary particles at high \pT. The correlated part of the uncertainty was removed. To estimate this systematic uncertainty, generator-level to reconstructed-level matchings in simulation with different requirements were considered. The main contribution comes from additional constraints on the transverse momentum of reconstructed track, with respect to the generated transverse momentum of the particle to which the track is matched. The uncorrelated part of this uncertainty is below 1\% up to 50\,\GeV\ and then rapidly increases with increasing \pT. At the highest measured \pT, it reaches 20\% for \PbPb and 15\% for \pp yields. 

The uncertainty associated with the unfolding reflects the need to choose a specific number of iterations to use in the unfolding procedure. It is estimated by changing the number of iterations, and observing the effect on the charged-particle yields. The uncertainty remains less than 1\% at low \pT\ and increases at high \pT\ where it reaches 8\%.

The uncertainty on the \pT\ resolution reflects possible differences in the distortion of track $\pT$ in MC simulation and data. To estimate this uncertainty, the \pT\ resolution (the width of the $1 - p_{\mathrm{T}}^\mathrm{gen}/p_{\mathrm{T}}^\mathrm{rec}$ distribution) was increased to 1.1 times the nominal width. The 10\% broadening was estimated from the resolution of muon \pT\ in the data. The uncertainty is 1\% at low \pT, but reaches 20\% at very high \pT.

The systematic uncertainty on the efficiency correction has several contributions. The first is the difference in the distributions for different $\eta$ ranges and both charges before merging, as explained in section~\ref{sec:correction}. This uncertainty is not higher than 2\%. 
The second contribution is from the uncertainty on the particle composition, estimated from comparisons of the relative rates of different identified particles types between simulation and the data measured by the ALICE collaboration~\cite{alice_particles}. This uncertainty is at most 3\%. Both corrections play an important role only at low \pT. The third uncertainty comes from the uncertainty associated with smoothing the track reconstruction efficiency correction at high \pT. This reaches 2\% at the highest \pT. The more significant uncertainty at high \pT\ comes from the uncertainty associated with the cross sections of the MC samples. These were varied, which resulted in an uncertainty of 5\% at the highest \pT\ due to the different acceptance and efficiency corrections of the MC samples.

The systematic uncertainty associated with possible mismodelling of the detector material is 2\% for $|\eta|<1$ and reaches 6\% in the highest $|\eta|$ region \cite{ATLASminb}.

\hspace*{1cm}

In general, the uncertainties associated with individual sources do not exceed 10\%. The exceptions are the track momentum resolution, on the fake tracks at high \pT\ and on the calculation of \Taa.  

\begin{table}[hb]
\begin{center}
\begin{tabular}{|l|ccccl|} \hline
	\multicolumn{6}{|c|}{Systematic uncertainties [\%]} \\ \hline
	       & \multicolumn{2}{c}{Spectra}  & $\rcp$       &  \rpbpb       & Strongest \\      
	Source & \PbPb & \pp &   &  & variation \\ \hline
	Luminosity &  & 3 & & 3 & \\
	\avgTaa  &  &  &  \multicolumn{2}{r}{1.5--13} & centrality\\	
	$\avgTaa/\avgTaaPerif$  &  &   \multicolumn{3}{c}{3.8--12}  & centrality \\	
	Jet trigger efficiency  & 1 & 3 & 1 & 3 & $\pT$ \\ 
	Track selection  & 10 & 4 & 10 & 10 & $\pT$ \\
	Fake and secondary tracks & 5 & 0.5 & 5 & 5 & $\pT$, centrality \\	
	Matching gen -- rec & 20 & 15 & 15 & 13 & $\pT$ \\
	Unfolding & 8 & 2 & 4 & 2 & $\pT$ \\
	\pT\ resolution & 20 & 7 & 14 & 12 & $\pT$ \\
	Efficiency correction & 5 & 1 & 4 & 4 & $\pT$, $\eta$ \\
	Detector material  & 2--6 & 2--6 &  &  & $\eta$ \\ \hline
\end{tabular}
\caption{Maximum values of systematic uncertainties in percent for the charged-particle spectra and the nuclear modification factors \rcp\ and \rpbpb. ``Fake and secondary tracks" reflect only the uncertainty at low \pT; the high-\pT\ part \mbox{is included in ``Matching gen -- rec".}}
\label{table:sysErr}

\end{center}
\end{table}

\section{Results} 
\label{sec:results}

\begin{figure}[t]
\centering
\includegraphics[width=0.7\columnwidth]{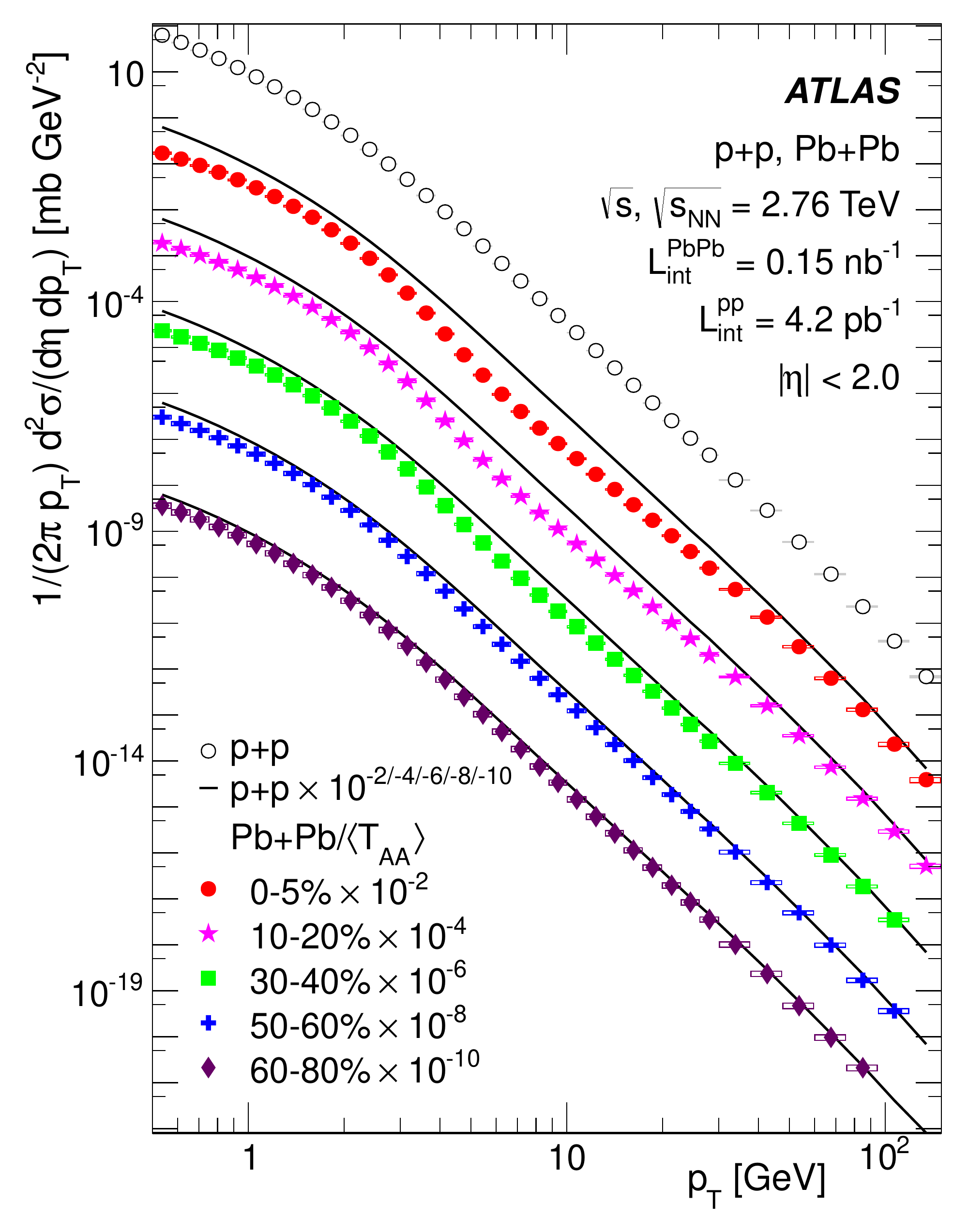}
\caption{Fully corrected charged-particle transverse momentum spectra for $|\eta|<2$ shown with filled markers in five centrality intervals: 0--5\%, 10--20\%, 30--40\%, 50--60\% and 60--80\% as well as the fully corrected charged-particle transverse momentum \pp\ cross section shown by open circles. Statistical uncertainties are smaller than the marker size. Systematic uncertainties are shown by open boxes. The different centrality intervals are scaled down by powers of ten for plot clarity. Each centrality interval is divided by the corresponding \avgTaa and plotted together with the \pp\ cross section scaled by the same factor shown with solid lines. The total systematic uncertainty on the \PbPb\ spectra includes the uncertainty of \avgTaa.}
\label{fig:spectra}
\end{figure}

\begin{figure}[t]
\centering
\includegraphics[width=0.7\columnwidth]{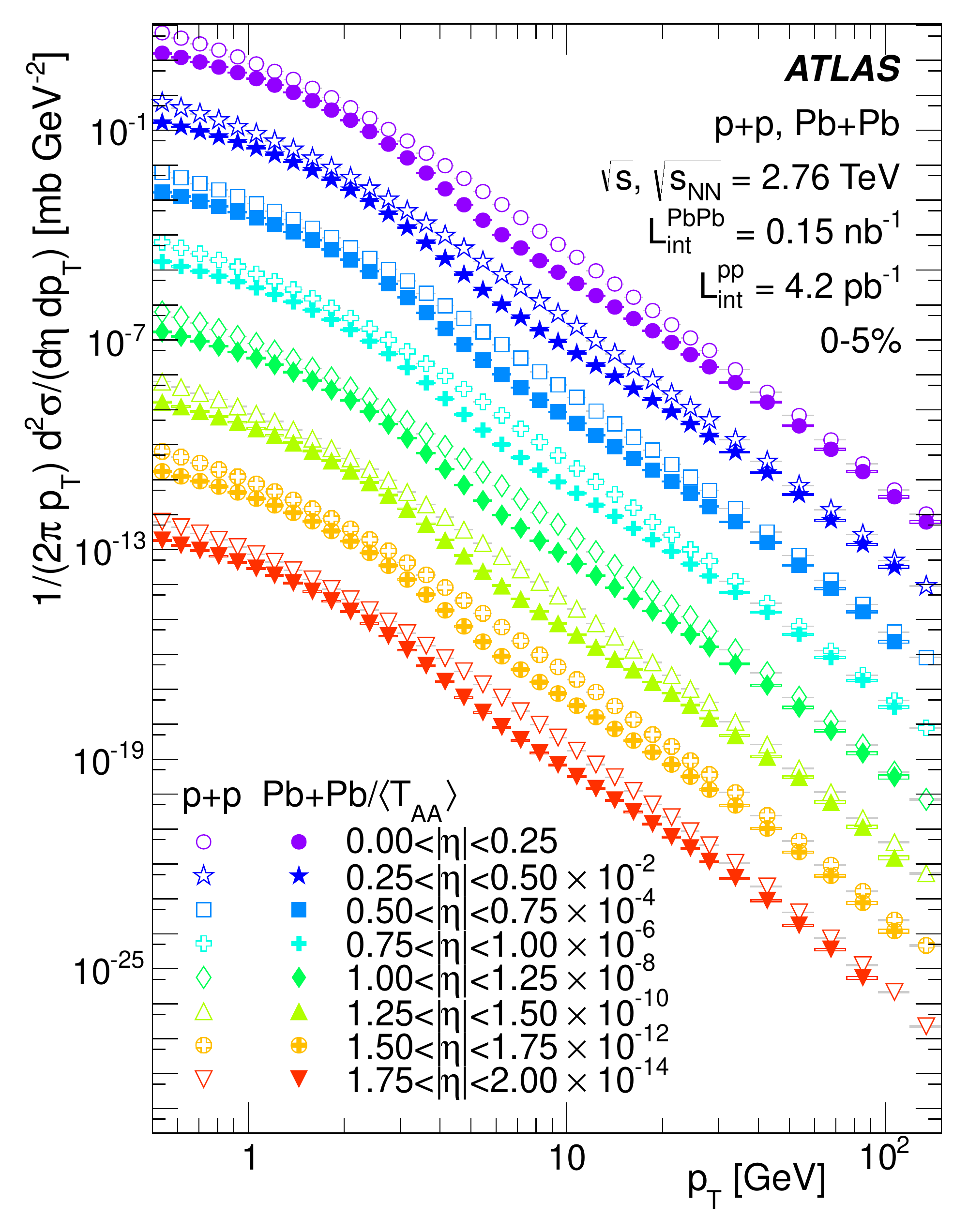}
\caption{Fully corrected charged-particle transverse momentum spectra produced in eight $|\eta|$ ranges shown with filled markers for the centrality interval 0--5\%. Statistical uncertainties are smaller than the marker size. Systematic uncertainties are shown with open boxes. The distributions in different $|\eta|$ intervals are scaled down by powers of ten for plot clarity. \PbPb\ spectra are divided by their corresponding \avgTaa and plotted together with the cross section measured in \pp\ collisions. These results are shown with open markers of the same style. The total systematic uncertainty of the \PbPb\ spectra includes the uncertainty of \avgTaa.}
\label{fig:spectra_with_eta_bins}
\end{figure}

The corrected charged-particle spectra measured in \PbPb collisions at \energy are shown in figure~\ref{fig:spectra} for the pseudorapidity range $|\eta|<2$ and for five centrality intervals: 0--5\%, 10--20\%, 30--40\%, 50--60\% and 60--80\% in the \pT\ range 0.5--150\,\GeV. In figure~\ref{fig:spectra_with_eta_bins}, charged-particle \PbPb\ spectra in the 0--5\% centrality interval are shown for eight regions of $|\eta|$. Both figures show the spectra divided by the \avgTaa of the corresponding centrality interval compared with the charged-particle production cross sections measured in \pp\ collisions at $\sqs=2.76$\,\TeV. 
	
The charged-hadron yields in peripheral \PbPb collisions, depicted by diamond markers in figure~\ref{fig:spectra}, show a \pT\ dependence similar to that of \pp\ collisions. Going from peripheral to central collisions, the $\Taa$-scaled \PbPb yields increasingly deviate from the \pp spectra. This deviation is largest for \pT\ less than 1\,\GeV\ and in the $\pT$ range 3--30\,\GeV.

\begin{figure}[ht]
\centering
\includegraphics[width=0.7\columnwidth]{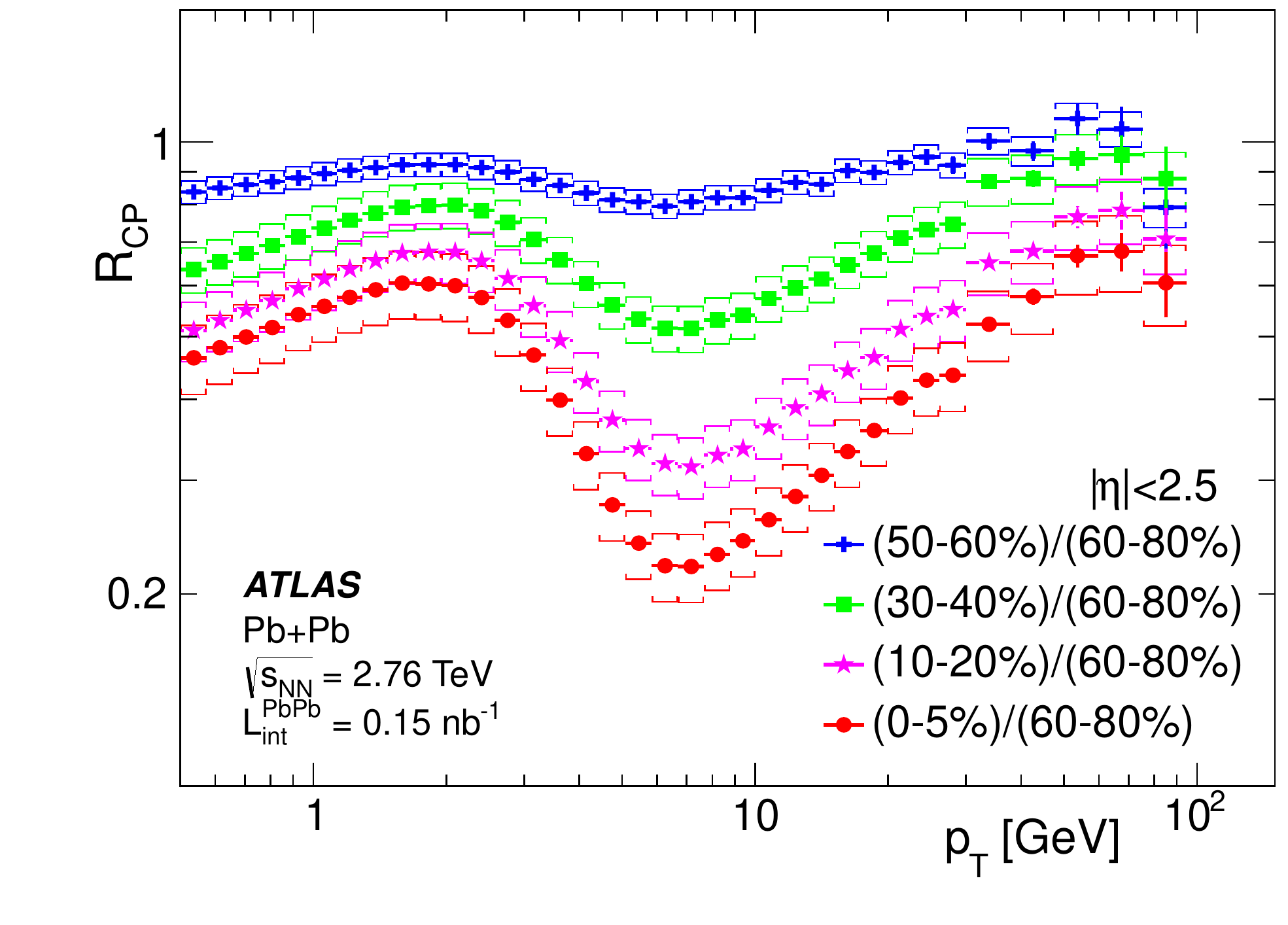}
\caption{The dependence of the central-to-peripheral nuclear modification factor \rcp on \pT\ measured in the centrality intervals 0--5\%, 10--20\%, 30--40\% and 50--60\%. The centrality class \mbox{60--80\%} forms the denominator. Statistical uncertainties are shown with vertical bars and systematic uncertainties with brackets.}
\label{fig:rcp}
\end{figure}

\begin{figure}[ht]
\centering
\includegraphics[width=0.7\columnwidth]{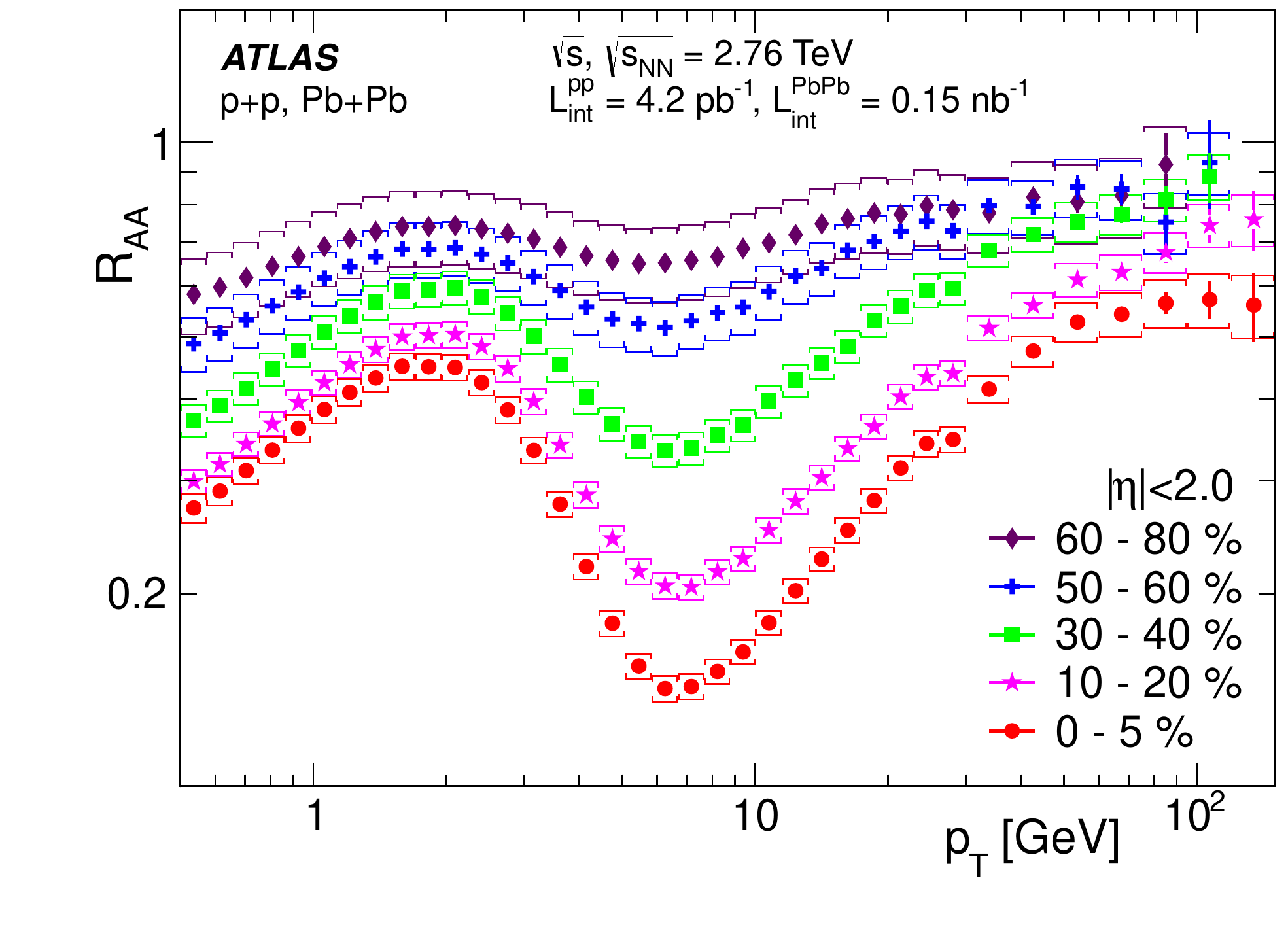}
\caption{The dependence of the nuclear modification factor \raa on \pT\ measured in the centrality intervals 0--5\%, 10--20\%, 30--40\%, 50--60\% and 60--80\%. Statistical uncertainties are shown with vertical bars and systematic uncertainties with brackets.}
\label{fig:raa}
\end{figure}

\begin{figure}[ht]
\centering
\includegraphics[width=0.7\columnwidth]{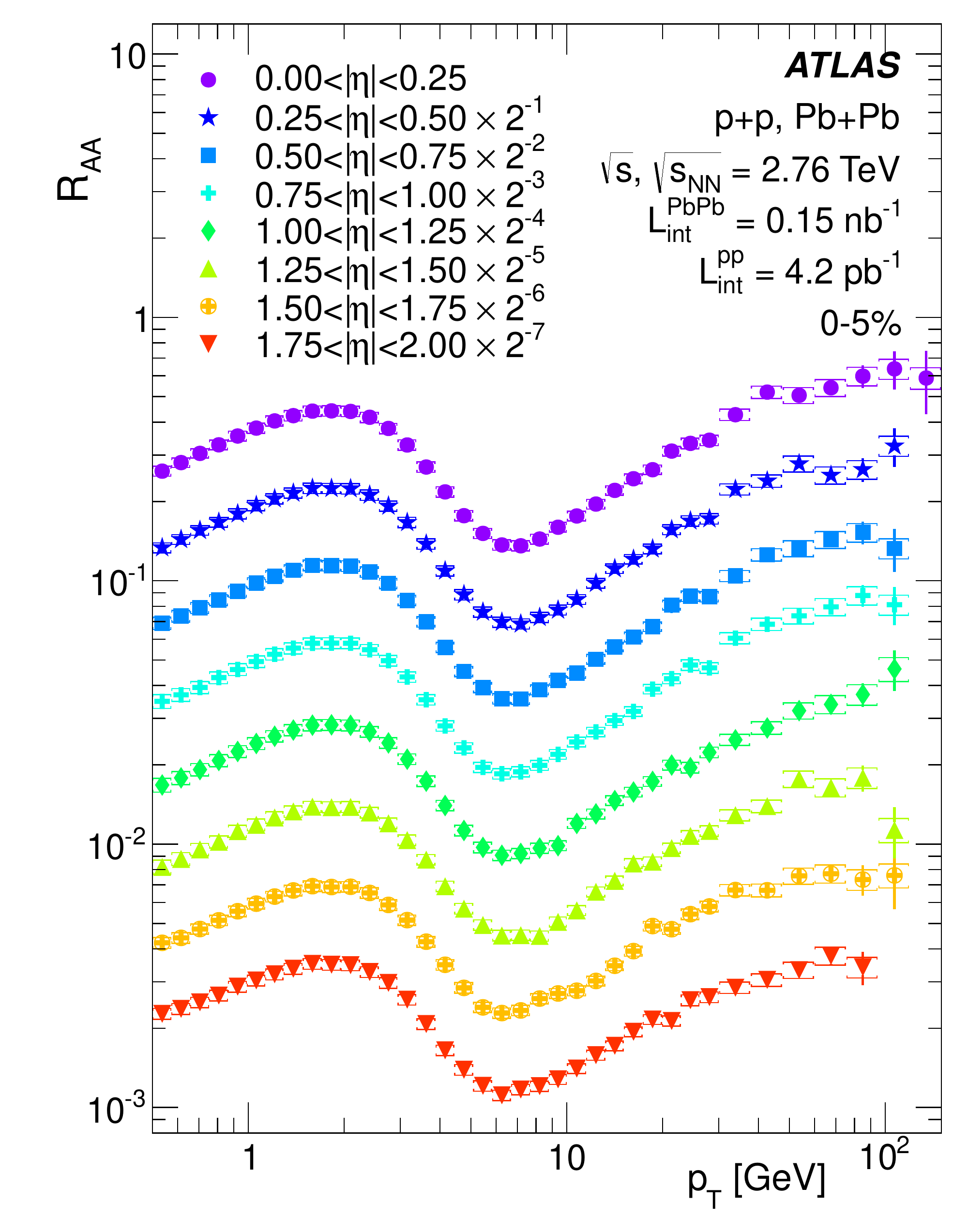}
\caption{The \raa dependence on \pT\ measured in eight $|\eta|$ ranges for the centrality interval \mbox{0--5\%.} Statistical uncertainties are shown with vertical bars and systematic uncertainties with brackets. The different $|\eta|$ ranges are scaled down by powers of two for plot clarity.}
\label{fig:raa_with_eta_bins}
\end{figure}

Figure~\ref{fig:rcp} shows the nuclear modification factor \rcp for four centrality classes (0--5\%, 10--20\%, 30--40\%, 50--60\%) with respect to the 60--80\% class. The \rcp\ as a function of \pT\ reaches a minimum of $0.22\pm0.03(syst.)$ at $\pT\approx 7$\,\GeV\ in the 0--5\% centrality class. The statistical uncertainty of the value at the minimum is negligible. The measurement of \rcp\ is restricted to $\pT<95$\,\GeV\ because it is limited by the small sample size in the 60--80\% centrality class.

\begin{figure}[!t]
\centering
\includegraphics[width=0.7\columnwidth]{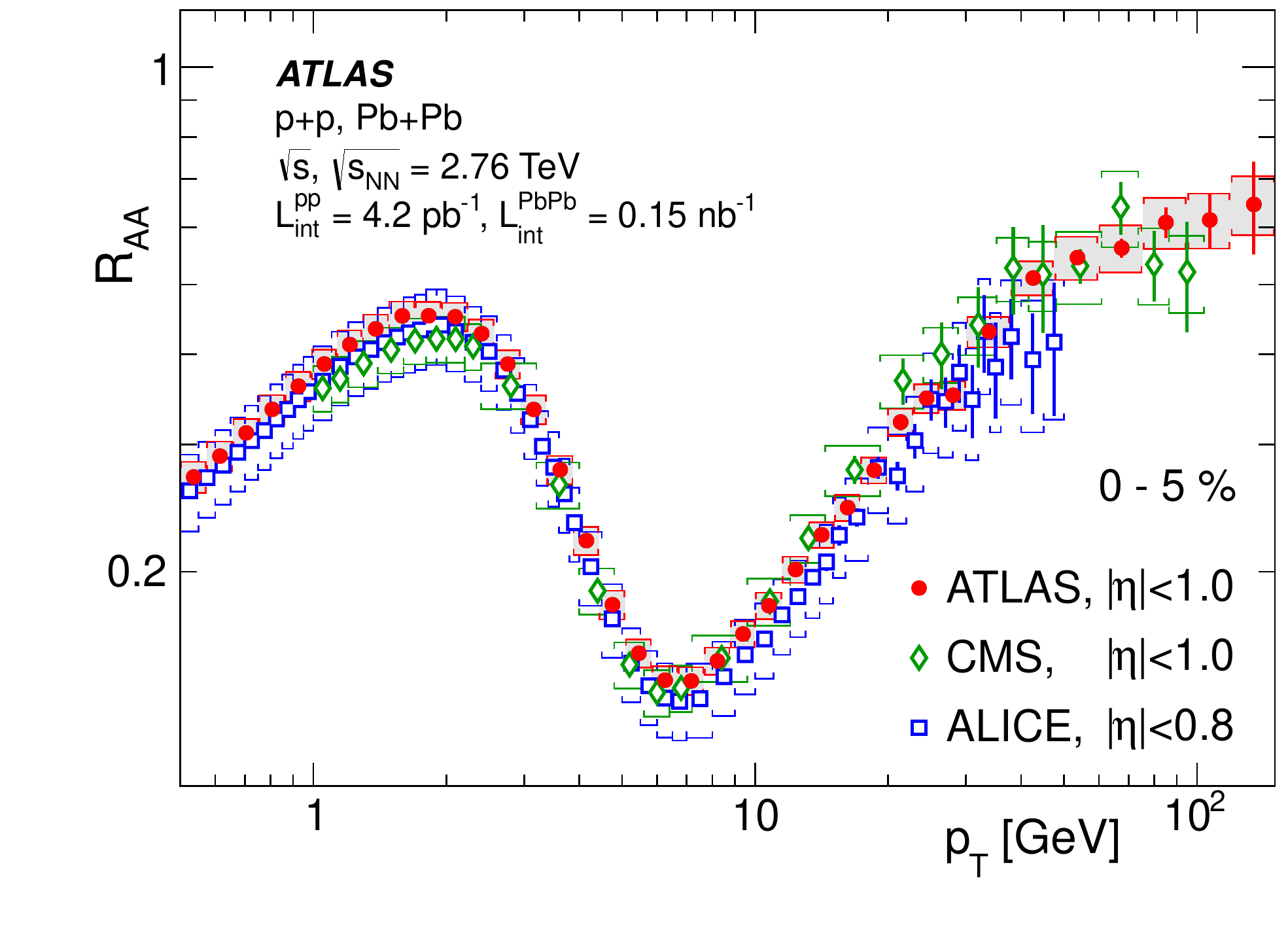}
\caption{The \raa dependence on \pT\ measured in the centrality interval 0--5\%. The ATLAS (shown as full circles) and CMS results~\cite{cms_raa} (shown as open diamonds) correspond to the pseudorapidity range $|\eta|<1$, the ALICE results~\cite{alice_raa} (open squares) to $|\eta|<0.8$. Statistical uncertainties are shown with vertical bars and systematic uncertainties with brackets.}
\label{fig:raa_comp}
\end{figure}

Figure~\ref{fig:raa} shows the nuclear modification factor \raa as a function of $\pT$ in five centrality intervals: 0--5\%, 10--20\%, 30--40\%, 50--60\% and 60--80\% constructed as the ratio of the \PbPb to \pp\ distributions from figure~\ref{fig:spectra}. Figure~\ref{fig:raa_with_eta_bins} shows \raa for 0--5\% collisions in eight $|\eta|$ ranges constructed as the ratio of the \PbPb to \pp\ distributions from figure~\ref{fig:spectra_with_eta_bins}. The \raa measured as a function of \pT\ shows a characteristic non-flat \pT\ shape which becomes more pronounced for more central collisions. They first increase with increasing \pT\ reaching a maximum at $\pT\approx 2$\,GeV, a feature commonly associated with the Cronin effect~\cite{CroninEffect}. Then, $\raa$ decreases with higher \pt\ reaching a minimum at $\pT\approx7$\,\GeV, where the charged-particle suppression is strongest. The rate of charged particles is noticeably suppressed even in the 60--80\% centrality interval, reaching a minimum of $0.66\pm0.09(syst.)$, where the effects of jet quenching are expected to be smallest. The suppression is strongest in the most central 0--5\% collisions, where the minimum is $0.14\pm0.01(syst.)$. The statistical uncertainties in this \pT\ region are negligible compared to the systematic uncertainties. Above this \pT, \raa generally increases with increasing \pT\ up to $\approx 60$\,\GeV\ and then reaches a plateau, consistent with zero slope, at a value of $0.55\pm0.01(stat.)\pm0.04(syst.)$ for the most central collisions. 

The comparison of the \raa\ values with previous measurements from CMS~\cite{cms_raa} and ALICE~\cite{alice_raa} in the most central (0--5\%) interval, common to all three experiments, is shown in figure~\ref{fig:raa_comp}. For this comparison the ATLAS \raa\ values are evaluated in the pseudorapidity interval $|\eta|<1$, the interval used by the CMS experiment. The ALICE experiment used a similar pseudorapidity interval, $|\eta|<0.8$. The comparison reveals agreement of the measurements within their uncertainties. In other centrality intervals common to ATLAS and ALICE, the results agree as well. The ATLAS results extend the \pT\ reach and precision of the previous measurements.

\begin{figure}[h]
\centering
\includegraphics[width=0.7\columnwidth]{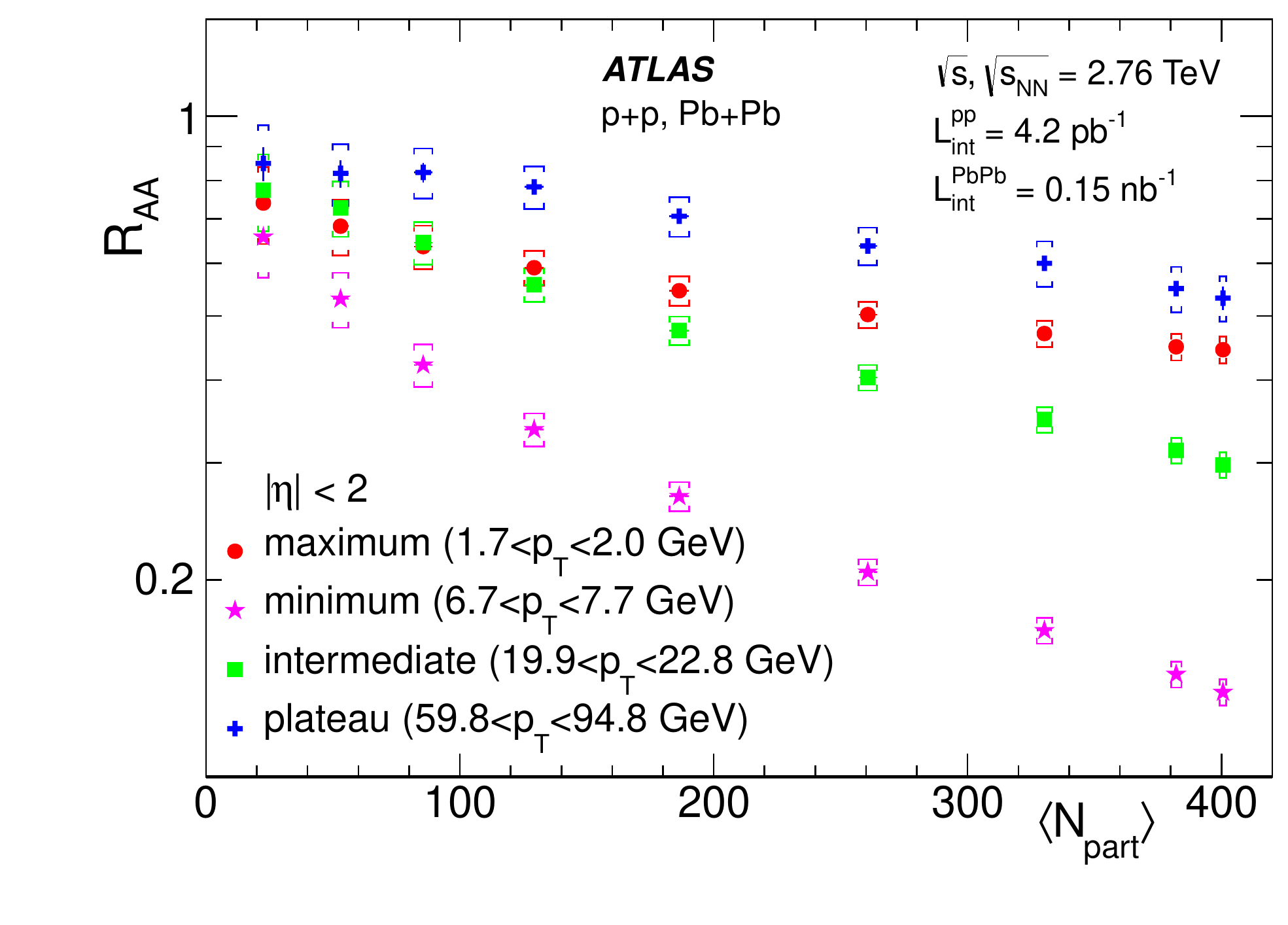}
\caption{The \raa values plotted as a function of \avgNpart measured in four transverse momentum intervals corresponding to the {\it maximum} \rpbpb ($1.7<\pT<2.0$\,\GeV), {\it minimum} \rpbpb \mbox{($6.7<\pT<7.7$\,\GeV)}, {\it intermediate} \rpbpb ($19.9<\pT<22.8$\,\GeV) and {\it plateau} of \rpbpb \mbox{($59.8<\pT<94.8$\,\GeV)}. The statistical uncertainties, which are smaller than the size of markers in all but a few cases, are shown with vertical bars and the systematic uncertainties with brackets.}
\label{fig:raa_cent}
\end{figure}

Figure~\ref{fig:raa_cent} shows \raa as a function of the mean number of participating nucleons, \avgNpart, in the four momentum intervals corresponding to the local {\it maximum} of \raa in the region $1.7<\pT<2.0$\,\GeV, the local {\it minimum} in the region $6.7<\pT<7.7$\,\GeV, the {\it plateau} region $59.8<\pT<94.8$\,\GeV\ and to the \pT\ interval where \raa has an {\it intermediate} magnitude, between the minimum and plateau value, in the region $19.9<\pT<22.8$\,\GeV. In all momentum intervals \raa decreases with \avgNpart , however the decrease is strongest for the {\it minimum} interval and weakest in the {\it plateau} region. In the {\it maximum} and {\it intermediate} momentum ranges the \raa values are consistent for $\avgNpart<100$, but then deviate at larger $\avgNpart$. In the most central collisions, \raa in the {\it maximum} interval is comparable to that in the {\it plateau} \pT\ region. Comparing the {\it plateau} region to the \raa dependence on \avgNpart measured with the jets of comparable energy~\cite{atlas_jet_raa}, both distributions show similar trend. They have also similar values for peripheral collisions, but in the more central collisions the jet \raa reaches lower values.

\begin{figure}[ht]
\centering
\includegraphics[width=0.7\columnwidth]{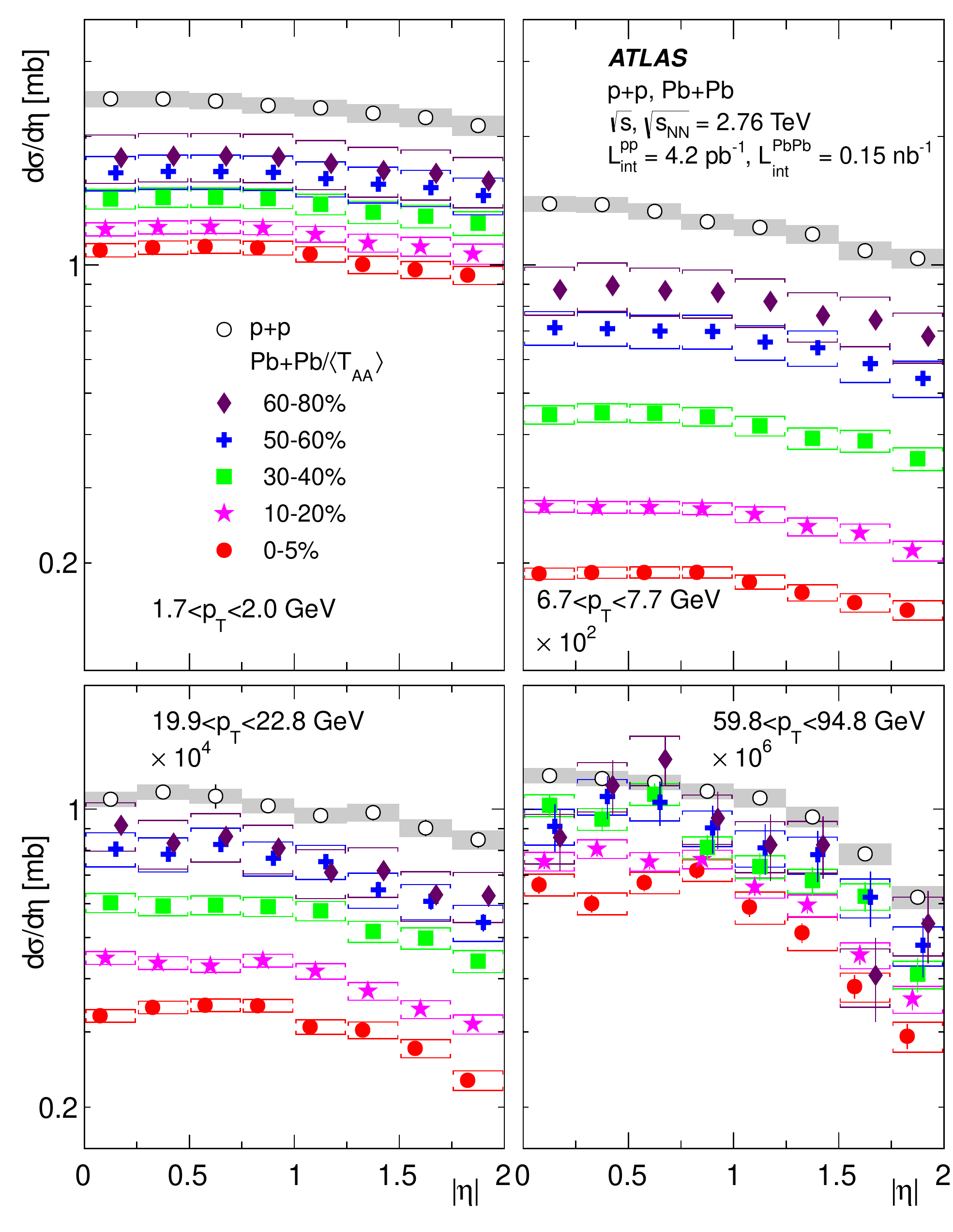}
\caption{Fully corrected charged-particle pseudorapidity distributions for four transverse momentum intervals defined in the text: {\it maximum}, {\it minimum}, {\it intermediate} and {\it plateau}. The different \pT\ ranges are scaled down by powers of ten. Markers are shifted from the centres of the bin for plot clarity. For each \pT\ range, data are shown in the centrality intervals: 0--5\%, 10--20\%, 30--40\%, 50--60\% and 60--80\%. The statistical uncertainties, which are typically smaller than the size of markers, are shown with vertical bars and the systematic uncertainties with brackets or grey bands. \PbPb\ spectra are divided by the corresponding \avgTaa and plotted together with the cross section measured in \pp\ collisions, shown with the open circles. The total systematic uncertainty of the \PbPb\ spectra includes the uncertainty of \avgTaa.}
\label{fig:rapidity}
\end{figure}

To further investigate the nature of the suppression mechanism, the pseudorapidity dependence of \raa was studied. Figure~\ref{fig:rapidity} shows the charged-particle cross section in \pp collisions and the $\Taa$-scaled yield in \PbPb collisions, \dsigmadeta, measured in the same centrality intervals as the spectra and \raa, integrated over the \pT\ intervals used to characterize the shape of \raa\ in figure~\ref{fig:raa_cent}. The \dsigmadeta distributions for \PbPb, divided by their corresponding \avgTaa, are compared to the \pp\ cross section. The shapes of the \PbPb \dsigmadeta distributions change slightly within the four specified $\pT$ ranges. At fixed \pT, the shapes agree for different centrality intervals as well as for the \pp data. 

\begin{figure}[ht]
\begin{center}
\includegraphics[width=0.7\columnwidth]{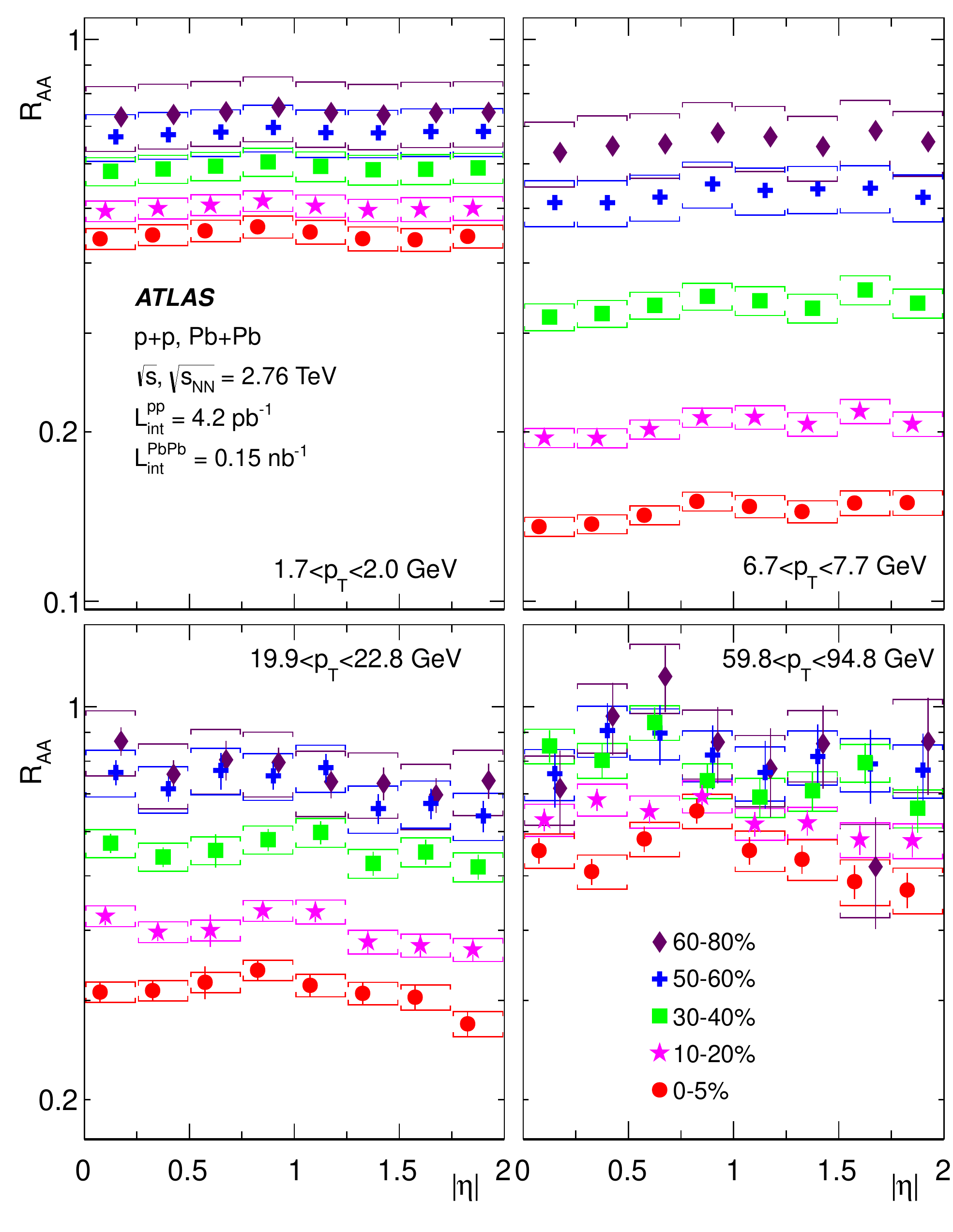}
\end{center}
\caption{The \raa dependence on $\eta$ in four transverse momentum intervals defined in the text: {\it maximum}, {\it minimum}, {\it intermediate} and {\it plateau}. For each \pT\ range, data are shown in centrality intervals: 0--5\%, 10--20\%, 30--40\%, 50--60\% and 60--80\%. Statistical uncertainties are shown with vertical bars and systematic uncertainties with brackets. Markers are shifted from the centres of the bin for plot clarity.}
\label{fig:raa_rap}
\end{figure}

Figure~\ref{fig:raa_rap} shows \raa measured in the four defined \pT\ intervals as a function of pseudorapidity. For all \pT\ intervals \raa shows little dependence on $|\eta|$ and is generally consistent with a flat behaviour within the statistical and systematic uncertainties. 

\clearpage
\section{Summary} 
\label{sec:conclusions}

A precise measurement of inclusive charged-hadron production in \PbPb collisions at \energy and in \pp collisions at \mbox{$\sqs=2.76$\,\TeV} is presented in the \pT\ interval from 0.5 to 150\,\GeV. The data were collected with the ATLAS detector at the LHC, and correspond to 0.15\,nb${}^{-1}$ of \PbPb collisions and 4.2\,pb${}^{-1}$ of \pp collisions. The results extend those of previous analyses, having a wider reach in \pT\ and increased precision, and they are important input to theoretical calculations.

In all centrality and rapidity intervals studied, the nuclear modification factor, \raa, shows a characteristic shape with a local maximum at $\pT \approx 2$\,\GeV\ decreasing to a minimum at $\pT \approx 7$\,\GeV. It increases again towards higher $\pT$ with a change in the slope at $\pT \approx 60$\,\GeV, where the rise becomes much less steep. Above this \pT, the \pT\ dependence is consistent with a flat behaviour within statistical and systematic uncertainties. This shape is present in all measured centrality and pseudorapidity intervals, but it has the largest maximum-to-minimum ratio in the most central collisions. The \rcp\ distribution follows a similar pattern, although the suppression is milder. 

For the 0--5\% most central collisions \raa reaches a value of $0.14\pm0.01(syst.)$ at the minimum. Then \raa\ increases with increasing \pT\ up to $\approx 60$\,\GeV. Above this \pT, \raa is consistent with reaching a plateau at the value of $\raa=0.55\pm0.01(stat.)\pm0.04(syst.)$. 

In the individual \pT\ intervals the shape of the charged-hadron pseudorapidity distribution is similar in all \PbPb centrality intervals and in the \pp data. At fixed \pT, the \raa values are consistent with a flat behaviour in $|\eta|$, within the statistical and systematic uncertainties, in the pseudorapidity range $|\eta|<2$. A similar flat behaviour was found for jet \raa as well~\cite{atlas_jet_raa}, although the \pT\ dependence is very different.

The \raa values decrease with increasing \avgNpart in all measured \pT\ intervals, but the $\avgNpart$ dependence is strongest around the minimum of \raa ($\pT\approx7$\,\GeV) and weakest in the $\pT$ region above 60\,\GeV.

\section*{Acknowledgments}
\label{sec:akn}

We thank CERN for the very successful operation of the LHC, as well as the
support staff from our institutions without whom ATLAS could not be
operated efficiently.

We acknowledge the support of ANPCyT, Argentina; YerPhI, Armenia; ARC,
Australia; BMWFW and FWF, Austria; ANAS, Azerbaijan; SSTC, Belarus; CNPq and FAPESP,
Brazil; NSERC, NRC and CFI, Canada; CERN; CONICYT, Chile; CAS, MOST and NSFC,
China; COLCIENCIAS, Colombia; MSMT CR, MPO CR and VSC CR, Czech Republic;
DNRF, DNSRC and Lundbeck Foundation, Denmark; EPLANET, ERC and NSRF, European Union;
IN2P3-CNRS, CEA-DSM/IRFU, France; GNSF, Georgia; BMBF, DFG, HGF, MPG and AvH
Foundation, Germany; GSRT and NSRF, Greece; RGC, Hong Kong SAR, China; ISF, MINERVA, GIF, I-CORE and Benoziyo Center, Israel; INFN, Italy; MEXT and JSPS, Japan; CNRST, Morocco; FOM and NWO, Netherlands; BRF and RCN, Norway; MNiSW and NCN, Poland; GRICES and FCT, Portugal; MNE/IFA, Romania; MES of Russia and NRC KI, Russian Federation; JINR; MSTD,
Serbia; MSSR, Slovakia; ARRS and MIZ\v{S}, Slovenia; DST/NRF, South Africa;
MINECO, Spain; SRC and Wallenberg Foundation, Sweden; SER, SNSF and Cantons of
Bern and Geneva, Switzerland; NSC, Taiwan; TAEK, Turkey; STFC, the Royal
Society and Leverhulme Trust, United Kingdom; DOE and NSF, United States of
America.

The crucial computing support from all WLCG partners is acknowledged
gratefully, in particular from CERN and the ATLAS Tier-1 facilities at
TRIUMF (Canada), NDGF (Denmark, Norway, Sweden), CC-IN2P3 (France),
KIT/GridKA (Germany), INFN-CNAF (Italy), NL-T1 (Netherlands), PIC (Spain),
ASGC (Taiwan), RAL (UK) and BNL (USA) and in the Tier-2 facilities
worldwide.

\bibliography{RPbPb_paper_2014}
\bibliographystyle{JHEP}

\clearpage


\begin{flushleft}
{\Large The ATLAS Collaboration}

\bigskip

G.~Aad$^{\rm 85}$,
B.~Abbott$^{\rm 113}$,
J.~Abdallah$^{\rm 152}$,
S.~Abdel~Khalek$^{\rm 117}$,
O.~Abdinov$^{\rm 11}$,
R.~Aben$^{\rm 107}$,
B.~Abi$^{\rm 114}$,
M.~Abolins$^{\rm 90}$,
O.S.~AbouZeid$^{\rm 159}$,
H.~Abramowicz$^{\rm 154}$
% ...
\bigskip
\\
% ...
$^{11}$ Institute of Physics, Azerbaijan Academy of Sciences, Baku, Azerbaijan\\
% ...
$^{85}$ CPPM, Aix-Marseille Universit{\'e} and CNRS/IN2P3, Marseille, France\\
% ...
$^{90}$ Department of Physics and Astronomy, Michigan State University, East Lansing MI, United States of America\\
% ...
$^{107}$ Nikhef National Institute for Subatomic Physics and University of Amsterdam, Amsterdam, Netherlands\\
% ...
$^{113}$ Homer L. Dodge Department of Physics and Astronomy, University of Oklahoma, Norman OK, United States of America\\
$^{114}$ Department of Physics, Oklahoma State University, Stillwater OK, United States of America\\
% ...
$^{117}$ LAL, Universit{\'e} Paris-Sud and CNRS/IN2P3, Orsay, France\\
% ...
$^{152}$ Institute of Physics, Academia Sinica, Taipei, Taiwan\\
% ...
$^{154}$ Raymond and Beverly Sackler School of Physics and Astronomy, Tel Aviv University, Tel Aviv, Israel\\
% ...
$^{159}$ Department of Physics, University of Toronto, Toronto ON, Canada\\
% ...
\end{flushleft}

\end{document}